\documentclass[acmtog]{acmart}

\newcommand{\final}{1}

\usepackage{algpseudocode}
\usepackage{graphicx}
\usepackage{xcolor}
\usepackage{acro}
\usepackage{xspace}
\usepackage{multirow}
\usepackage{subcaption}
\usepackage{listings}
\usepackage{wrapfig}
\usepackage[ruled,vlined]{algorithm2e}

\usepackage{import}
\usepackage{xifthen}
\usepackage{pdfpages}
\usepackage{transparent}
\usepackage{placeins}

\usepackage{tabularx}
\newcolumntype{Y}{>{\raggedright\arraybackslash}X}
\usepackage{array}       %
\usepackage{multirow}    %
\usepackage{booktabs}    %

\usepackage{minted}
\usepackage[most]{tcolorbox}
\tcbuselibrary{listings,minted,breakable}

\usepackage{tikz}
\usetikzlibrary{spy}

\definecolor{ListBGColor}{rgb}{0.95,0.95,0.95}
\definecolor{ListCommentColor}{rgb}{0.3,0.5,0.3}
\definecolor{ListKeywordColor}{rgb}{0.2,0.2,0.7}
\definecolor{ListStringColor}{rgb}{0.6,0.1,0.1}
\definecolor{ListNumberColor}{rgb}{0.4,0.4,0.4}
\definecolor{ListIdentifierColor}{rgb}{0.1,0.1,0.1}

\newtcblisting{cppcode}[2][]{%
  listing engine=minted,
  minted language=c++,
  breakable,
  enhanced,
  colback=black!1,
  colframe=black!16,
  boxrule=0.6pt,
  arc=2mm,
  left=6mm, right=2mm, top=1.5mm, bottom=1.5mm,
  listing only,
  minted options={
      fontsize=\small,
      linenos,
      breaklines,
      autogobble,
      tabsize=2,
      escapeinside=||,
      numbersep=6pt,
      style=friendly,%
      texcomments=true
    },
  title={#2},
  #1
}

\newcommand*{\Cpp}{C\texttt{++}}
\definecolor{linkblue}{RGB}{0,92,175}

\definecolor{ahmedColor}{rgb}{1,0.43,0.43}
\newcommand{\ahmed}[1]{{\color{ahmedColor} ahmed: #1}}
\newcommand{\justin}[1]{\textbf{\scriptsize{\color{blue}JS: #1}}}

\newcommand{\rahul}[1]{\textbf{\scriptsize{\color{orange}RG: #1}}}

\definecolor{todoColor}{rgb}{0, 0, 1}
\newcommand{\todo}[1]{{\color{todoColor} todo: #1}}

\definecolor{changeColor}{rgb}{0,0.45,0}

\newcommand{\warning}[1]{{\emph{\color{red} #1}}}
\newcommand{\note}[1]{{\emph{\color{blue} #1}}}

\newcommand{\nothing}[1]{}

\ifthenelse{\equal{\final}{1}}
{
\renewcommand{\ahmed}[1]{}
\renewcommand{\rahul}[1]{}
\renewcommand{\justin}[1]{}
\renewcommand{\todo}[1]{}
\renewcommand{\warning}[1]{}
\renewcommand{\note}[1]{}

}
{}

\definecolor{grayColor}{rgb}{0.5,0.5,0.5}

\usepackage{enumitem}
\setlist[enumerate]{noitemsep, nolistsep, leftmargin=*}
\setlist[itemize]{noitemsep, nolistsep, leftmargin=*}

\setcopyright{cc}
\setcctype{by}
\acmJournal{TOG}
\acmYear{2026} 
\acmVolume{45} 
\acmNumber{4} 
\acmArticle{50}
\acmMonth{7} 
\acmDOI{10.1145/3811338}

\acmSubmissionID{744}

\begin{document}

\title{Locality-Aware Automatic Differentiation on the GPU for Mesh-Based Computations}

\author{Ahmed H.\ Mahmoud}
\affiliation{%
  \department{Computer Science \& Artificial Intelligence Laboratory}
  \institution{Massachusetts Institute of Technology}
  \streetaddress{32 Vassar St}
  \city{Cambridge}
  \state{MA}
  \postcode{02139}
  \country{USA}
  }  
\email{ahdhn@mit.edu}
\orcid{0000-0003-1857-913X}

\author{Rahul Goel}
\affiliation{%
  \department{Computer Science \& Artificial Intelligence Laboratory}
  \institution{Massachusetts Institute of Technology}
  \streetaddress{32 Vassar St}
  \city{Cambridge}
  \state{MA}
  \postcode{02139}
  \country{USA}
  }  
\email{goel@.csail.mit.edu}
\orcid{0000-0002-9564-4022}

\author{Jonathan Ragan-Kelley}
\affiliation{%
  \department{Computer Science \& Artificial Intelligence Laboratory}
  \institution{Massachusetts Institute of Technology}
  \streetaddress{32 Vassar St}
  \city{Cambridge}
  \state{MA}
  \postcode{02139}
  \country{USA}
  }  
\email{jrk@mit.edu}
\orcid{0000-0001-6243-9543}

\author{Justin Solomon}
\affiliation{%
  \department{Computer Science \& Artificial Intelligence Laboratory}
  \institution{Massachusetts Institute of Technology}
  \streetaddress{32 Vassar St}
  \city{Cambridge}
  \state{MA}
  \postcode{02139}
  \country{USA}
  }  
\email{jsolomon@mit.edu}
\orcid{0000-0002-7701-7586}

\begin{teaserfigure}
  \centering
  \includegraphics[width=\textwidth]{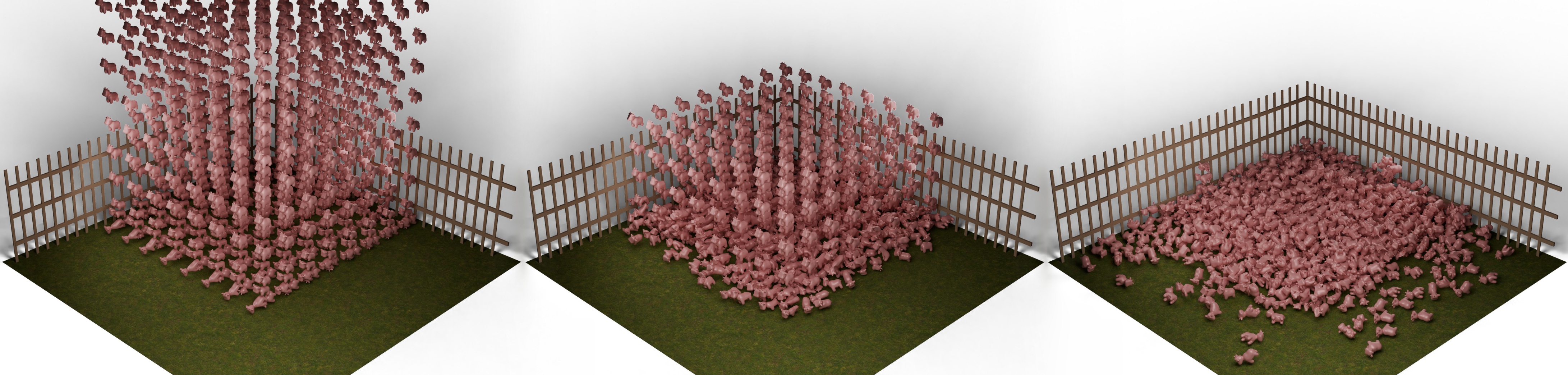}
    \caption{We introduce a GPU system for efficient automatic differentiation of computations defined on triangle meshes that exploits locality and sparsity in mesh-based workloads. Using our system, users specify only the energy terms of their application while our system computes gradients, sparse Hessians, and Jacobians automatically and efficiently on the GPU\@. Here, we use a Newton solver for large-scale elastic shell simulation to simulate $\approx$ 700 Spot cows ($\approx$2.1M vertices in total) falling to the ground where derivative computation accounts for only 12.2\% of the total runtime.}
  \label{fig:teaser}
\end{teaserfigure}


\begin{abstract}
	We present a GPU-based system for automatic differentiation (AD) of functions defined on triangle meshes, designed to exploit the locality and sparsity in mesh-based computation. Our system evaluates derivatives using per-element forward-mode AD, confining all computation to registers and shared memory and assembling global gradients, sparse Jacobians, and sparse Hessians directly on the GPU\@. By avoiding global computation graphs, intermediate buffers, and device-host synchronization, our approach minimizes memory traffic and enables efficient differentiation under both static and dynamically changing sparsity. 
	Our programming model lets users express energy terms over mesh neighborhoods, while our system automatically manages parallel execution, derivative propagation, sparse assembly, and matrix-free operations such as Hessian-vector products. Our system supports both scalar- and vector-valued objectives, dynamic interaction-driven sparsity updates, and seamless integration with external GPU sparse linear solvers. 
	We evaluate our system on applications including elastic and cloth simulation, surface parameterization, mesh smoothing, frame field design, ARAP deformation, and spherical manifold optimization. Across these tasks, our system consistently outperforms state-of-the-art differentiation frameworks, including PyTorch, JAX, Warp, Dr.JIT, EnzymeAD, and Thallo. We demonstrate speedups across a range of solver types, from Newton and Gauss-Newton for nonlinear least squares to L-BFGS and gradient descent, and across different derivative usage modes, including Hessian-vector products as well as full sparse Hessian and Jacobian construction. Our system is available as open source at \href{https://github.com/owensgroup/RXMesh}{\textcolor{linkblue}{https://github.com/owensgroup/RXMesh}}.
\end{abstract}

\maketitle

\begin{figure}    
    \captionsetup[subfloat]{labelformat=empty}
	\subfloat[]{\includegraphics[width=0.22\textwidth]{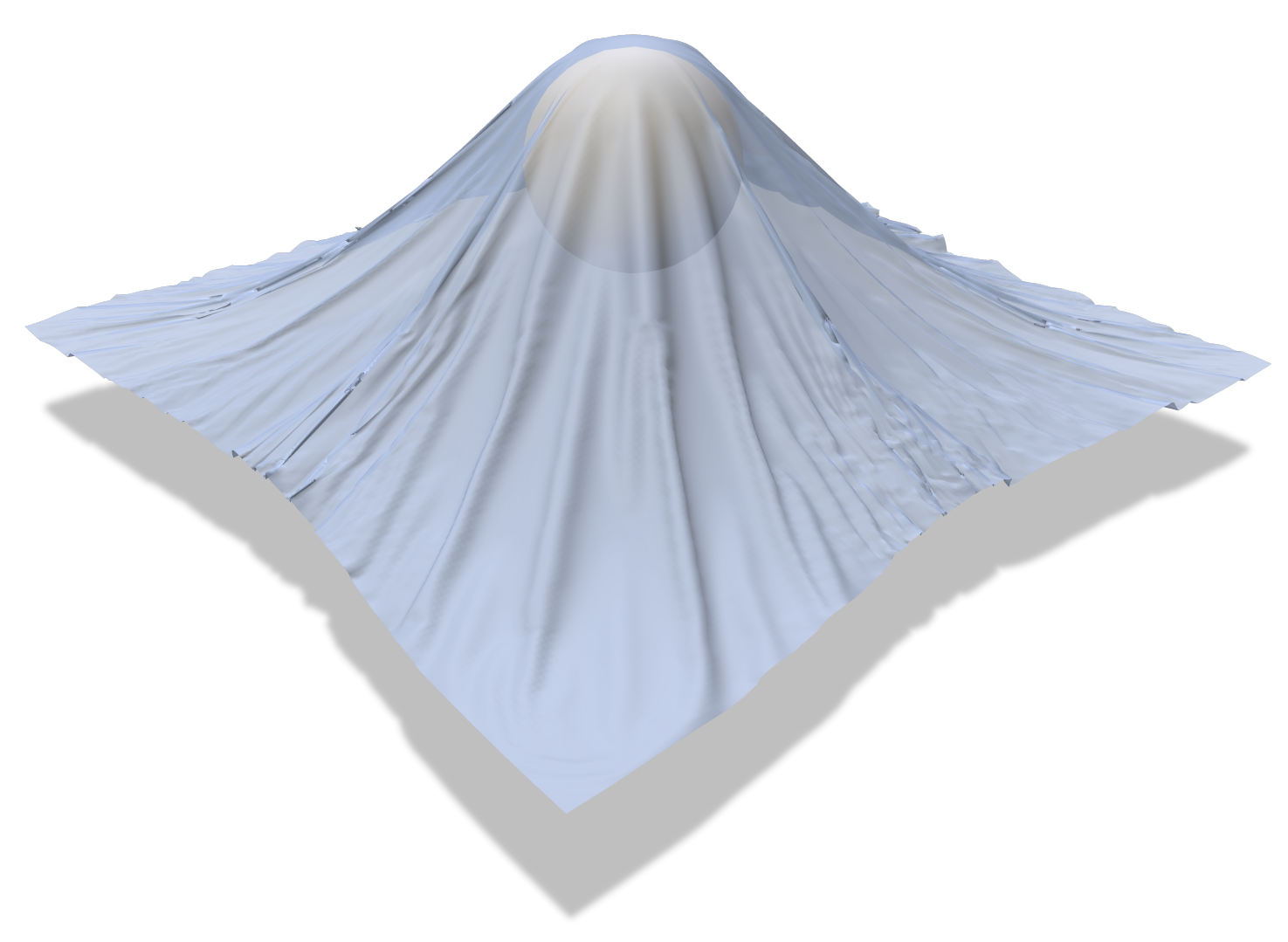}}
	\subfloat[]{ \begin{tikzpicture}[spy using outlines={rectangle, magnification=5, size=0.8cm, connect spies}]
        \node {\includegraphics[width=0.24\textwidth]{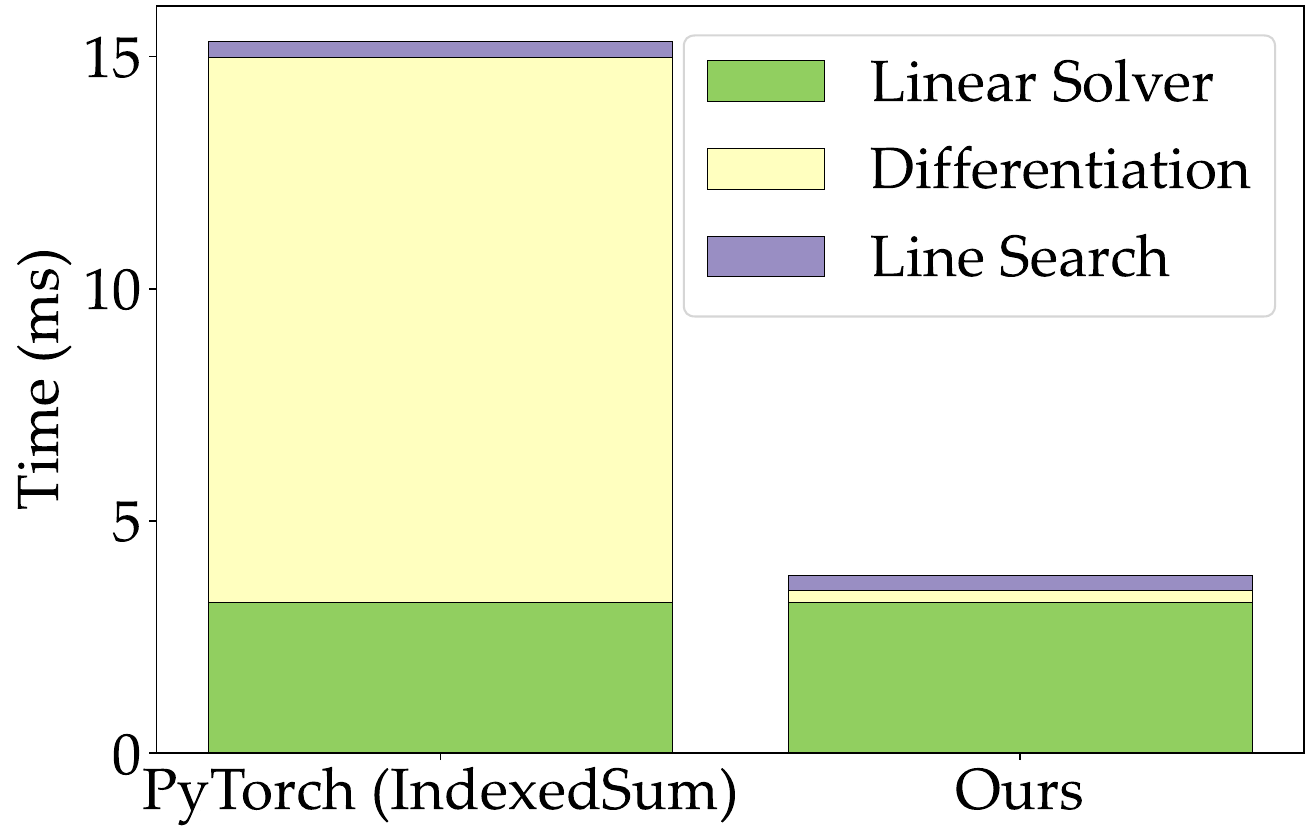}};
        \spy on (0.8, -0.58) in node [right] at (1.2,-0.1);
    \end{tikzpicture}}
    \caption{Prior to our system, nonlinear optimization using Newton's method (shown here as the time for a single Newton iteration of a mass--spring cloth simulation on a $100^2$-vertex mesh) was bottlenecked by differentiation. Our system significantly accelerates derivative computation, shifting the bottleneck to the linear solver.}{}
    \label{fig:ms_breakdown}
\end{figure}

\section{Introduction}
\label{sec:intro}

Countless algorithms for science and engineering applications rely on a common building block of evaluating the derivatives of a functional defined over a mesh. Derivatives computation most often arises in optimization-driven tasks in physical simulation, inverse problems, design automation, and geometry processing, where gradients and higher-order derivatives directly determine convergence, stability, and overall runtime. In many of these settings, derivative evaluation dominates the computational cost, making its efficiency a first-order concern rather than just an implementation detail.

The widespread adoption of (stochastic) gradient-based methods in machine learning has spurred the development of highly optimized automatic differentiation (AD) frameworks. Much of this progress has been guided by the needs of machine learning workloads where the dominant computational model consists of \emph{dense tensors} and large, regular computation graphs. Systems such as PyTorch~\cite{Ansel:2024:P2F} and JAX~\cite{Bradbury:2018:JCT} excel in this regime, enabling end-to-end differentiation of neural networks.

In contrast, scientific and graphics workloads, particularly those involving meshes, operate in a different regime. Mesh-based problems often involve large, sparse systems with localized dependencies induced by irregular connectivity (see Figure~\ref{fig:mesh_ad}). As resolution increases to meet accuracy requirements in applications such as biomechanics, structural engineering, and fluid simulation, the resulting Jacobians and Hessians grow in size while remaining sparse.

General-purpose AD frameworks fail to exploit this structure since they represent derivatives through dense or implicitly dense intermediate forms, resulting into poor runtime performance and excessive memory usage when applied to mesh-based problems. In practice, derivative evaluation can become a dominant cost, in some cases exceeding the cost of the downstream linear solve itself (see Figure~\ref{fig:ms_breakdown}). This inefficiency forces practitioners to limit problem size, avoid second-order methods, or resort to hand-derived derivatives to achieve acceptable performance~\cite{Huang:2024:DSF}.

Mesh-based applications therefore require AD systems that preserve sparsity by construction and support both efficient first- and second-order derivatives computation without constraining the choice of numerical solver. While prior mesh-oriented AD tools exist (see \S\ref{sec:mesh_tools}), they typically target a narrow derivative order, assume specific solver structures, or fail to scale efficiently to large systems.

AD in mesh-based computations on the GPU is memory-bound, i.e., the primary cost lies not in arithmetic operations but in accessing and updating large, sparse data structures. Modern GPUs provide extremely high bandwidth and low latency in registers and shared memory but only when computation is organized to exploit \emph{locality}. Conventional AD systems struggle to do so where their derivative representations induce scattered reads and writes to global memory, preventing effective use of the GPU memory hierarchy. As a result, much of the available bandwidth remains unused.

In this paper, we present a GPU AD system tailored to the sparsity and locality of computation on triangle meshes. Our system exploits this locality to perform differentiation at the level of individual mesh elements. Each gradient, Jacobian, or Hessian contribution depends only on a small, fixed neighborhood, which allows us to keep differentiation entirely within registers or shared memory and limit global memory traffic (Figure~\ref{fig:assemble}).

Our system supports efficient computation of first- and second-order derivatives, i.e., gradients, sparse Jacobians, sparse Hessians, and Hessian--vector products. We use forward-mode AD via operator overloading to evaluate per-element derivatives independently, assemble sparse derivatives directly on the GPU, and optionally operate in matrix-free mode. In addition to fixed mesh stencils, the system supports dynamic pairwise interactions that introduce new couplings between elements during execution, requiring updates to the Hessian sparsity pattern. By preallocating sparse structures from mesh connectivity and managing sparsity updates explicitly on the GPU, we avoid dynamic computation graph construction and enable fully parallel execution under both static and dynamically evolving sparsity. Our implementation is open source and available at \href{https://github.com/owensgroup/RXMesh}{\textcolor{linkblue}{https://github.com/owensgroup/RXMesh}}.

With a focus on triangle meshes, we design a system for sparse differentiation on meshes to achieve the following design goals:
\begin{enumerate}
    \item \textbf{Performance:} Achieve high performance for the core operations in mesh-based optimization pipelines that require sparse differentiation, including assembling sparse Hessians and Jacobians, evaluating Hessian--vector products, and handling dynamically changing sparsity patterns.
    \item \textbf{Robustness:} Support a broad range of mesh-based operations and energy formulations, including terms defined on vertices, edges, faces, and dynamically generated interactions (e.g., between different bodies), as well as first- and second-order derivatives. The system places no restrictions on mesh quality, accommodating non-manifold meshes and disconnected components.
    \item \textbf{Interoperability:} Enable easy integration with existing high-performance numerical GPU solvers as well as accelerated spatial data structures (e.g., BVHs). 
    \item \textbf{Decoupling specification from execution:} Allow users to specify \emph{what} local computations define their objective independently of \emph{how} these computations are evaluated, differentiated, and assembled via an simple programming interface. A single problem specification can be reused across multiple evaluation modes (passive evaluation, full differentiation, and matrix-free derivative products) and across different solver pipelines, without modification.
\end{enumerate}

We test our system on a suite of applications including mass-spring cloth simulation, mesh parameterization, manifold optimization, mesh smoothing, ARAP deformation, curl-free polyvector design, and elastic simulation. We compare against widely used general-purpose AD frameworks, including PyTorch, JAX, Warp, EnzymeAD, and Dr.JIT, as well as the domain-specific nonlinear least-squares system Thallo~\cite{Mara:2021:TSF}. Across all benchmarks, our system consistently achieves higher performance for different workloads that involve sparse first- and second-order derivatives.

In summary, this paper presents a GPU system for automatic differentiation on triangle meshes that preserves sparsity and exploits locality by evaluating derivatives at the level of local mesh neighborhoods. The system supports first- and second-order derivatives, explicit sparse gradient/Jacobian/Hessian construction, Hessian-vector products, and dynamic sparsity updates arising from runtime interactions, all within a unified programming model. Across a range of mesh-based optimization and simulation problems, we show that this design substantially reduces differentiation cost and outperforms existing AD frameworks and domain-specific baselines.

\section{Related Work}
\label{sec:related}

There are several classical and modern techniques for computing derivatives in scientific computing and optimization~\cite{Martins:2021:EDO}. Each method offers different trade-offs in terms of accuracy, efficiency, expressiveness, and implementation effort~\cite{Kim:2022:DDI, Kim:2020:DDI}.

\emph{Manual differentiation} relies on deriving and implementing gradient expressions by hand. While this can produce highly optimized code for fixed formulations, it is labor-intensive, error-prone, and difficult to maintain or extend as models evolve. \emph{Symbolic differentiation} constructs closed-form derivative expressions by manipulating algebraic representations of a program. This approach can yield exact derivatives and enables algebraic simplification, common subexpression elimination, and other global optimizations. However, it scales poorly to large programs and struggles with control flow, loops, and iterative solvers that are ubiquitous in scientific computing. As a result, symbolic methods are typically restricted to small expressions or serve as building blocks within hybrid systems, and their extension to more general programs remains an active area of research~\cite{Herholz:2022:SSC, Fernandez:2025:SEB, Herholz:2024:AMB}.

\emph{Finite differences} approximate derivatives using truncated Taylor expansions. They are simple to implement and applicable to black-box functions, but are sensitive to step-size selection, suffer from truncation and cancellation errors, and scale poorly with input dimensionality. \emph{Complex-step differentiation} addresses some numerical issues of finite differences by avoiding subtractive cancellation through complex arithmetic. While highly accurate, it still incurs a linear cost in the number of input variables. %

The focus of our work, \emph{automatic differentiation} (AD), also known as algorithmic differentiation, applies the chain rule directly to program execution and computes derivatives to machine precision. AD combines the generality of finite differences with the accuracy of symbolic methods, while remaining applicable to programs with complex control flow and iterative structure. These properties have made AD the dominant approach for differentiation in modern scientific and machine learning software.

We categorize existing AD systems into two broad groups. The first consists of general-purpose, domain-agnostic frameworks that can differentiate any computation expressed within their programming model. The second is domain-specific systems designed for geometric computation, and in particular for mesh-based workloads. We also provide an overview of RXMesh~\cite{Mahmoud:2021:RAG} since our system relies on its data structure.

\subsection{Automatic Differentiation (AD)}
\label{sec:ad}
We begin by briefly reviewing the two main modes of AD as they are central to the rest of the paper. For a more comprehensive treatment, we refer readers to the textbooks by \citet{Naumann:2012:TAO} and \citet{Griewank:2008:EDP}.

AD is a family of techniques to compute derivatives of functions expressed as computer programs. AD exploits the fact that any such program is a composition of elementary operations with known derivatives allowing for systematic application of the \emph{chain rule}. There are two primary modes of AD, i.e., \emph{forward} and \emph{reverse mode}.

In forward mode, derivatives propagate from inputs to outputs, computing the directional derivative of a function $f : \mathbb{R}^n \to \mathbb{R}^m$ along a chosen tangent direction. Forward mode maintains a derivative (or dual) value alongside each intermediate variable during program execution. A common implementation strategy is operator overloading where each arithmetic operation is redefined to also compute and propagate the derivative. This approach aligns naturally with the execution order of the original program where each node carries forward both its value and its local derivative.

In reverse mode, derivatives are propagated from outputs to inputs, computing the gradient of a scalar-valued function $f : \mathbb{R}^n \to \mathbb{R}$ by traversing the computation graph in reverse. Reverse mode requires first recording a computation graph during the forward pass which captures all intermediate variables and dependencies. During the reverse pass, this graph is traversed backward to apply the chain rule and accumulate gradients with respect to the inputs.

\subsection{AD Tools}
\label{sec:tools}

ADOL-C~\cite{Griewank:1996:A7A} is a foundational C/\Cpp\ library that computes gradients and higher-order derivatives using operator overloading and taping. While broadly applicable, its tape management introduces runtime overhead and limits compiler optimizations, making it less suited for performance-critical tasks. PyTorch~\cite{Ansel:2024:P2F} and JAX~\cite{Bradbury:2018:JCT} are more recent AD systems that are widely used in machine learning and numerical computing with efficient support for dense tensor operations. However, their execution models are less effective for sparse, irregular structures or fine-grained control flow typical in scientific computing and geometric data processing.

Enzyme~\cite{Moses:2021:RMA, Moses:2022:SAD} performs AD at the LLVM IR level, enabling compiler-level optimizations that reduce memory and execution overhead---especially in reverse mode. This makes it attractive for integrating AD into existing high-performance codebases. Enoki~\cite{Jakob:2019:ESV} is a \Cpp17 library that supports forward and reverse-mode AD with vectorized execution across CPU and GPU\@. Dr.JIT~\cite{Jakob:2022:DAJ}, its successor, compiles high-level Python/\Cpp\ code into optimized machine code, enabling efficient differentiable rendering. While effective in rendering contexts, both systems are limited to first-order derivatives and do not exploit the sparsity or locality common in mesh-based computations. Our system builds on the insights behind these tools, adapting them to better suit sparse workloads in mesh processing, leading to improved performance in our target domain.

\subsection{Mesh AD Tools}
\label{sec:mesh_tools}

\citet{Herholz:2022:SSC} proposed a system that applies symbolic differentiation to unoptimized \Cpp\ code operating on sparse data. Their approach constructs a global expression graph across the mesh, eliminates redundant subexpressions, and generates vectorized, parallel kernels for CPU or GPU execution. While this yields highly optimized code, it incurs high memory usage and long compilation times. To address these issues, \citet{Herholz:2024:AMB} introduced a refinement where users define symbolic expressions for a single mesh element. The system compiles these elementwise kernels independently of the global mesh, significantly reducing memory usage and compile times.

TinyAD~\cite{Schmidt:2022:TAD} is a \Cpp\ library designed for sparse optimization. It computes gradients and Hessians by differentiating small, per-element subproblems. Users define local energy terms and the system applies forward-mode AD to compute first- and second-order derivatives. While structurally similar to our method, TinyAD runs only on CPUs and uses OpenMP for parallelism. Our work builds on TinyAD's structure but is designed explicitly for GPUs by optimizing for memory locality and parallel throughput.

Opt~\cite{Devito:2017:OAD} is a DSL for nonlinear least squares problems. It uses symbolic differentiation at the IR level and generates first-order optimization code. Thallo~\cite{Mara:2021:TSF} extends Opt by improving scheduling and memory layout. Unlike our system, these frameworks focus on first-order optimization for nonlinear least squares problems and do not support explicit sparse Hessian construction or Hessian-vector products. Warp~\cite{Macklin:2022:WAH} is a Python-based framework for high-performance spatial computing. Users write simulation kernels in Python which are JIT-compiled for CPU or GPU\@. Warp supports reverse-mode AD by generating backward kernels enabling simulation differentiation and integration with ML pipelines.

\emph{Relation to \citet{Herholz:2024:AMB}:}
Our system and \citet{Herholz:2024:AMB} address similar challenges in sparse differentiation for mesh-based computation, including support for second-order derivatives and dynamically changing Hessian sparsity induced by interacting mesh elements. Both systems aim to achieve high performance, but differ in how this goal is pursued. \citet{Herholz:2024:AMB} is based on symbolic differentiation, constructing and optimizing global expression graphs to eliminate redundant computations before generating parallel code via their symbolic backend~\cite{Herholz:2022:SSC}. In contrast, our approach prioritizes memory locality and bandwidth efficiency, organizing differentiation to occur at the finest granularity possible and confining most computation to registers and shared memory on the GPU\@. While \citet{Herholz:2024:AMB} focuses primarily on expression-level optimization, our system emphasizes aggressive memory optimizations and ensures that all stages of differentiation, including collision handling and sparse Hessian updates, execute entirely on the GPU\@. The two approaches are complementary and our system could serve as a backend for symbolic frontends such as \citet{Herholz:2024:AMB}, combining expression-level optimization with GPU-efficient sparse differentiation.

In summary, while several mesh-based AD tools support GPU execution, many overlook the importance of memory locality. Some focus solely on symbolic simplification~\cite{Herholz:2022:SSC,Herholz:2024:AMB}, or lack higher-order derivatives (e.g., Opt, Thallo, Warp), and some run only on the CPU~\cite{Schmidt:2022:TAD}. In contrast, our method treats the GPU as a first-class target and optimizes all AD operations with an emphasis on memory locality and execution efficiency.

\subsection{RXMesh Overview}
\label{sec:rxmesh}
RXMesh~\cite{Mahmoud:2021:RAG, Mahmoud:2025:DMP} is a GPU system for triangle mesh processing that supports both static meshes and dynamic workloads that modify connectivity at runtime. Its core idea is to partition the mesh into small \emph{patches} sized to fit the GPU memory hierarchy well so that most local computation can be carried out from fast on-chip shared memory rather than global memory. To preserve locality near patch boundaries, RXMesh augments each patch with ghost elements, called \emph{ribbons}, that cache neighboring out-of-patch mesh data. Each patch is encoded independently using compact sparse matrix representations of face-edge and edge-vertex incidence. Computation is organized at the patch level, with one CUDA block assigned to a patch so that threads cooperate on local queries with reduced divergence and improved load balance.

A ribbon element is treated as an indirection to its owner patch rather than as a locally stored element. Any access to its connectivity or attributes is resolved by first identifying the owner patch and the element's local index within that patch, and then reading the data from the owner's storage. This indirection is realized through a hash tables that store enough information to recover the ribbon element's owner patch and its local index within that patch. Any subsequent connectivity or attribute access to the ribbon element is then forwarded to the owner's storage.

\begin{figure}
    \includegraphics[width=0.49\textwidth]{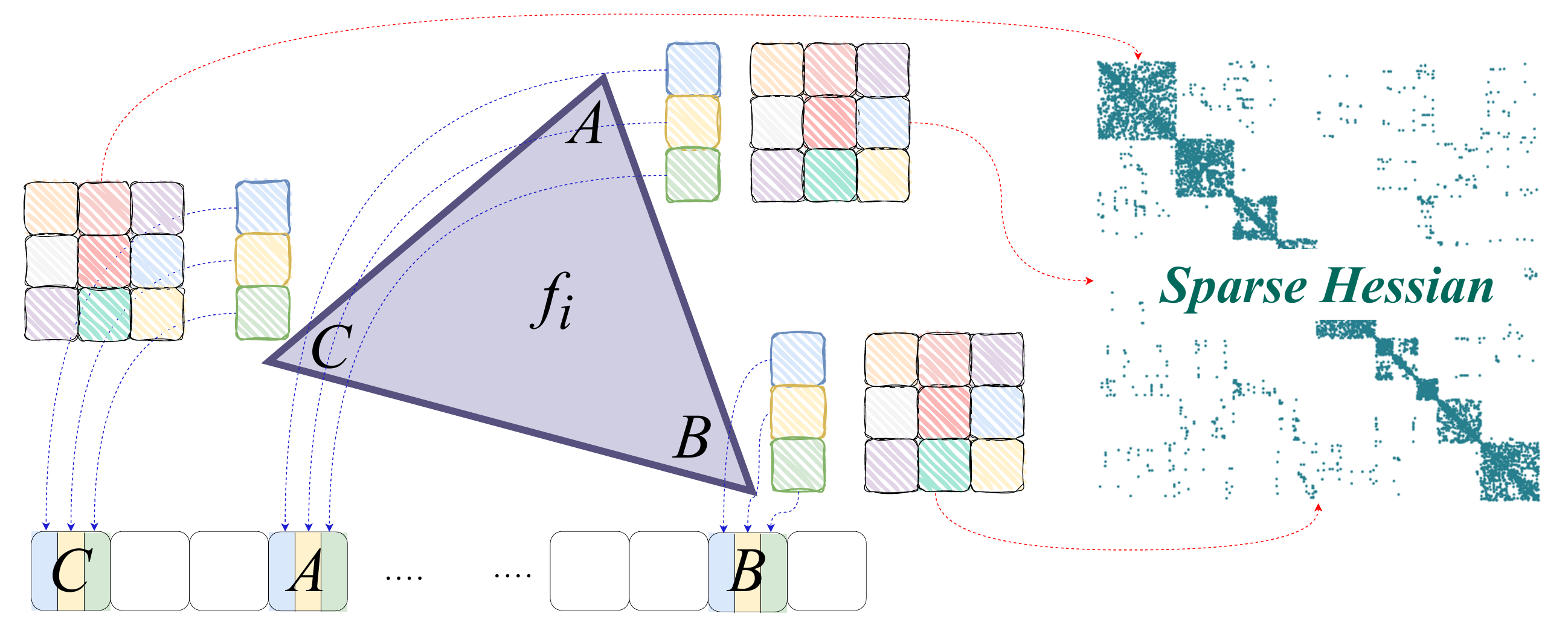}
    \caption{Per-element derivative computation and assembly. For a local energy term $f_i$ over triangle $ABC$, we compute derivatives with respect to local variables ($\in \mathbb{R}^3$) using compact indexing. We evaluate gradients and Hessians in registers/shared memory as small dense vectors/matrices via forward-mode AD, then we map them to global indices and accumulate them into the global pre-allocated gradient, sparse Hessian, or sparse Jacobian.}{}
    \label{fig:assemble}
\end{figure}

\section{System Overview}
\label{sec:overview}
At a high level, our system provides a way for the user to define mesh-based objectives by composing many local terms while the system takes responsibility for evaluating, differentiating, and assembling these terms efficiently on the GPU\@. From the user's perspective, an objective is built incrementally by specifying local energy or constraint contributions over mesh elements or element neighborhoods. These terms may depend on mesh adjacency information or on dynamically generated interactions, e.g., proximity-based or collision-driven couplings. The user expresses only the local computation, which then can be evaluated passively, differentiated to produce gradients and sparse Hessians/Jacobian, or used in matrix-free form for Hessian-vector products.

The novelty of our system lies in how this specification is realized into efficient GPU programs. While most of the computations required for derivative evaluation are \emph{local} to individual mesh elements, existing AD systems force users into an uncomfortable choice: (1)~rely on general-purpose frameworks that fail to preserve sparsity and therefore make second-order derivatives expensive, or (2)~hand-derive gradients and Hessians tailored to a specific problem.

\textbf{To bridge the gap between local problem descriptions and high-performance implementations, our system transforms the user's specification into an efficient GPU code.} Our system analyzes the structure of each local term to determine its stencil, dimensionality, and derivative requirements. We use this information to define how local computations map to global degrees of freedom and to preallocate the global data structures needed for sparse gradients, Jacobians, and Hessians. At runtime, local terms are evaluated independently and in parallel across the mesh. Each term computes its contribution using a compact local state, produces dense local derivatives, and contributes these results to global sparse structures (Figure~\ref{fig:assemble}), or applies them directly in matrix-free form. When interactions introduce new couplings during execution, the system updates sparsity patterns on the GPU without rebuilding global computation graphs or CPU--GPU transferring data\@.

Internally, we build our system around \emph{patch-level} execution, where the mesh is partitioned into small, independent patches that can be processed cooperatively within a GPU thread block. Each patch fits in shared memory and provides a bounded working set for both primal evaluation and differentiation. This allows all intermediates required for forward-mode AD, including temporary dual values and local Jacobians or Hessians, to be generated and consumed locally without spilling to global memory. We build on RXMesh's~\cite{Mahmoud:2021:RAG} patch-based execution model to achieve this locality but extend it substantially to support automatic differentiation, dynamic interaction terms, and sparse derivative assembly. While patch-based execution has been explored for efficient mesh processing~\cite{Mahmoud:2025:DMP,Chang:2022:MAC}, our work is the first to use patches for AD, demonstrating that patch locality a key factor of efficient sparse derivative computation on GPUs\@.

\section{Programming Model}
\label{sec:prog_model}
We target %
mesh-based problems in which the objective decomposes into a sum of \emph{local} functions defined over mesh elements. Such problems arise in simulation, optimization, and geometric data processing, where energies, constraints, or residuals are associated with vertices, edges, faces, or small neighborhoods thereof. A key property of these problems is that they are \emph{partially separable}~\cite{Nocedal:2006:NO}, i.e., each local term depends only on a small subset of the global degrees of freedom.

\subsection{Problem Overview}

\paragraph{Scalar-valued objectives.}
We first consider scalar-valued energy function $F:\mathbb{R}^n \rightarrow \mathbb{R}$, of the form
\begin{equation}
    F(x) = \sum_{j \in \mathbb{E}} f_j(x_j),
    \label{eq:scalar_energy}
\end{equation}
where $\mathbb{E}$ denotes a set of mesh elements (e.g., vertices, edges, or faces), and each $f_j : \mathbb{R}^{k_j} \rightarrow \mathbb{R}$ is a localized energy term associated with element $j$. Here, $x \in \mathbb{R}^n$ encodes the global degrees of freedom, while $x_j \in \mathbb{R}^{k_j}$ collects only those variables that influence $f_j$. The size $k_j$ is determined by the element's local neighborhood and is typically small and bounded (e.g., three vertices for a face, two for an edge).

This local dependency can be expressed using a binary \emph{selection matrix} $S_j \in \{0,1\}^{k_j \times n}$ such that $x_j = S_j x.$ Differentiation distributes over summation, allowing each local term to be differentiated independently. Specifically, we compute per-element gradients $g_j \in \mathbb{R}^{k_j}$ and Hessians
$H_j \in \mathbb{R}^{k_j \times k_j}$, which are assembled into global structures via
\begin{equation}
    g = \sum_{j \in \mathbb{E}} S_j^\top g_j
    \qquad \text{and} \qquad
    H = \sum_{j \in \mathbb{E}} S_j^\top H_j S_j .
    \label{eq:grad}
\end{equation}

\paragraph{Vector-valued functions and sparse Jacobians.}
In addition to scalar objectives, our programming model also supports vector-valued functions $F:\mathbb{R}^n \rightarrow \mathbb{R}^m$, which commonly arise in nonlinear least-squares problems and constraint formulations. We assume $F$ decomposes as a sum of local vector-valued terms, as in Equation~\ref{eq:scalar_energy}. Here, each $f_j:\mathbb{R}^{k_j} \rightarrow \mathbb{R}^{m_j}$ produces a small vector-valued local contribution associated with element $j$. The total output dimension is $m = \sum_j m_j$.

Differentiating $F$ yields a sparse Jacobian $J \in \mathbb{R}^{m \times n}$. Each local term contributes a dense local Jacobian $J_j \in \mathbb{R}^{m_j \times k_j}$, which is assembled into the global Jacobian via
\begin{equation}
    J = \sum_{j \in \mathbb{E}} P_j^\top J_j S_j,
    \label{eq:jac}
\end{equation}
where $P_j \in \{0,1\}^{m_j \times m}$ is a binary selection matrix that maps the $m_j$ local residuals of term $j$ to their locations in the global output vector. As in the scalar case, each $J_j$ is small and dense while the global Jacobian is sparse with a structure determined by mesh connectivity.

\begin{figure}
    \centering
    \begin{cppcode}[]{}

        using T = float;
        Mesh mesh ("input.obj");
        constexpr int VarDim = 3;
        Problem<T, VarDim, VertexHandle> problem(mesh);|\label{line:problem}|

        // Energy term
        problem.add_term<Op::EV>( |\label{line:op}|
        [=](EdgeHandle  eh,     //Edge Handle
        VertexIterator  iter,   //Edge's vertices
        VertexAttribute var) {  //Optimization variables
            //The edge's vertex position as Active variables
            auto x0 = var.active<ActiveT, 3>(eh, iter, 0); |\label{line:lift}|
            auto x1 = var.active<ActiveT, 3>(eh, iter, 1);
            //Local energy implementation, i.e., edge len
            ActiveT d = (x0 - x1).squaredNorm();
            return d;});

        problem.eval_terms(); |\label{line:eval}|
        T f = problem.get_current_energy();
        // Access the optimization variables (problem.var),
        // gradient (problem.grad),
        // sparse Hessian (problem.hess),
        // and sparse Jacobian (problem.jac)


    \end{cppcode}
    \captionof{lstlisting}{Example use of our programming model. The user defines a local energy term over mesh elements (in this case, per edge), specifies the required neighborhood access (\texttt{Op::EV} indicates that for each edge \texttt{E}, the term accesses its vertices \texttt{V}), and relies on the system to handle all parallel evaluation and differentiation. The global energy, gradient, and Hessian are assembled automatically from local contributions \protect\footnotemark .}
    \label{list:api}
\end{figure}

\subsection{Formulating Mesh-Based Objectives.}
Our programming model is inspired by RXMesh~\cite{Mahmoud:2021:RAG}, TinyAD~\cite{Schmidt:2022:TAD}, and \citet{Herholz:2024:AMB}. Users express objectives by writing local energy or residual functions over mesh elements while the system automatically manages derivative tracking, memory layout, and parallel execution.

Listing~\ref{list:api} illustrates the typical user workflow in our system. After loading the input mesh, the user specifies the floating-point type, the mesh element on which optimization variables are defined (vertices in this example), and the per-element variable dimension \texttt{VarDim} (Line~\ref{line:problem}).

The user then defines a local energy term over a specific mesh stencil. Local terms are expressed as \emph{lambda functions} that compute the energy or vector-valued contribution for a single element. The inputs to the lambda include a handle to the element, an iterator over its local neighborhood, and the optimization variables (Line~\ref{line:op}). Computations that should participate in differentiation must be expressed using the \texttt{ActiveT} type, where variables declared as \texttt{ActiveT} are treated as differentiable, while constants or auxiliary values may use standard floating-point types. The lambda returns an \texttt{ActiveT} value. Active and passive variables can be freely mixed within the computation.

In Listing~\ref{list:api}, the local term is written in standard CUDA/\Cpp\ and thus supports the full range of language constructs that arise in practice. In particular, users may freely introduce branching and early exits, e.g., to implement piecewise energies or penalties. This enables \emph{conditional element evaluation} where the user can skip the corresponding computation when a term is inactive for a given element. More generally, local terms may also contain bounded iterative control flow, such as a small while loop used to compute or refine a local quantity. The following example shows a per-face term that computes a square root using Newton iteration.

\begin{cppcode}[]{}
    problem.add_term<Op::F>(
    [=](FaceHandle fh, FaceAttribute var) {
        // Positive face quantity
        ActiveT a = var.active<ActiveT>(fh);
        // Newton iteration for sqrt(a)
        ActiveT x = a;
        while (abs(x * x - a) > tol) {
            x = 0.5 * (x + a / x);
        }
        return x;});
\end{cppcode}

\footnotetext{For brevity, we omit the \texttt{\_\_device\_\_} annotation in code snippets. In the implementation, the lambda is compiled for device execution.}

The formulation also supports different \emph{stencils} and neighborhoods. A term may operate on a single mesh element or on a local neighborhood, expressed through the \emph{operator tag}. For example, we use \texttt{Op::V}, \texttt{Op::E}, \texttt{Op::F} for element-only terms (as shown below), or \texttt{Op::FV}, \texttt{Op::EV}, \texttt{Op::EF} for neighborhood-based stencils. This makes it straightforward to compose objectives that mix heterogeneous local structures, e.g., combining per-vertex external forces, per-triangle stretch, and per-edge bending in a single optimization problem. The code below shows the per-vertex gravitational potential used in the cloth simulation application (\S\ref{sec:spring}), including an optional conditional evaluation.

\begin{cppcode}[]{}
    problem.add_term<Op::V>(
    [=](VertexHandle vh, VertexAttribute var){
        // Optional conditional evaluation:
        if (is_fixed(vh)) return ActiveT(0);
        // Vertex position is the differentiable variable
        auto x = var.active<ActiveT, 3>(vh);
        // Gravitational potential:
        // E = m * g * z
        return mass(vh) * gravity * x[2];});
\end{cppcode}

Finally, the mesh can carry arbitrary user-defined attributes that participate in local computations. Beyond primary unknowns (e.g., vertex positions in the above code), users can attach additional fields, e.g., velocities, per-element material parameters, stiffness values, or region labels, and reference them directly inside local energy computation. Likewise, global parameters, e.g., a time-step size, penalty weights, or material constants can be declared as ordinary \Cpp\ variables and captured by the lambda function. This design also accommodates non-standard scalar types, e.g., including complex-valued active variables---which we use in the frame field design application (\S\ref{sec:polyvector}). This keeps objective definitions compact while allowing the full flexibility needed by simulation and geometry-processing workloads. %

We also support vector-valued local terms for residual-based formulations (e.g., nonlinear least squares and constrained objectives). The API is identical to the scalar case, except that the user specifies the per-element residual dimension $N$ as an additional template argument. The local term then returns an \texttt{ActiveT<N>} value representing an $N$-vector of active quantities, from which the system constructs and assembles the corresponding sparse Jacobian. For example, the following snippet adds a soft positional constraint on a subset of vertices, i.e., for each constrained vertex, we return the 3D displacement from its target position as a 3-vector residual.

\begin{cppcode}[]{}
    constexpr int N = 3;
    problem.add_term<Op::V, N>(
    [=](VertexHandle vh, VertexAttribute var) {
        if (!is_constrained(vh)) return ActiveT<N>(0);
        // current position (active)
        auto x  = var.active<ActiveT, 3>(vh);
        // target position (passive)
        auto xt = target_pos(vh);
        ActiveT<N> r = x - xt;
        return r;});
\end{cppcode}

Once defined, our system evaluates these local terms in parallel across the mesh. While users write these terms in standard \Cpp, we instantiate these lambda functions as device code and executed as CUDA kernels on the GPU\@. Depending on the evaluation mode, the system can execute terms either in an active context, producing derivatives, or in a passive context, evaluating only the objective value. In the active case, local contributions are assembled into the global objective, gradient, sparse Hessian, or sparse Jacobian. For residual-based formulations, the system additionally supports sum-of-squares objectives and matrix-free Hessian--vector products, enabling Newton-like solvers without explicitly forming global derivative matrices.

\subsection{Interaction Terms over Dynamically Generated Element Pairs}
\label{sec:interaction_terms}
In addition to mesh-local objectives defined over fixed neighborhoods, our programming model supports \emph{interaction terms} defined over dynamically generated pairs of mesh elements. Such terms arise in many applications, including collision avoidance, self-interaction, proximity-based penalties, and other non-adjacent couplings that cannot be expressed using static mesh connectivity alone. For example, in elastic simulation, collision handling generates interaction terms between pairs of vertices or faces that come within a prescribed distance, even though they are not adjacent in the mesh connectivity~\cite{Li:2020:IPC}.

An interaction term is a function defined over a \emph{pair of elements} where the set of interacting pairs is determined at runtime. Each pair induces a local contribution that depends on the degrees of freedom associated with both elements and therefore introduces off-diagonal blocks in the global Hessian or Jacobian. Despite this global coupling, each interaction term remains local in the sense that it depends only on a small, fixed number of variables.

\paragraph{Pair generation.}
Our programming model places no restrictions on how interacting element pairs are identified. Depending on the application, a pair may consist of two vertices, a face and a vertex, or other combinations of mesh elements. Users are free to employ any application-specific strategy to generate such pairs, including spatial acceleration structures, hash-based methods, or problem-specific predicates. The only requirement is that the resulting set of pairs be explicitly provided to the system. Pair generation is designed to be fully parallel, i.e., multiple GPU threads may insert pairs concurrently, enabling scalable construction of large, data-dependent interaction sets, as shown below.

\begin{cppcode}[]{}
    // Phase 1: generate an interaction set
    // Any GPU kernel may insert pairs concurrently
    // (e.g., using a BVH, hash grid, or predicates).
    __global__
    void build_pairs(Problem problem /*, ... */){
        // candidate vertices from a spatial query
        VertexHandle h0, h1;
        if (should_interact(h0, h1)) {
            // concurrent insertion
            problem.interaction_pairs.insert(h0, h1);}}
\end{cppcode}

\paragraph{Pairwise local terms.}
Once an interaction set has been constructed, users define pairwise local terms in the same spirit as mesh-local objectives. Each term is evaluated independently for every interaction pair and computes a local energy based on the associated elements. The API mirrors that of standard local terms, except that the iterator now ranges over user-provided element pairs rather than mesh-adjacency neighborhoods. The operator tag specifies the type of interaction, e.g., \texttt{Op::VV} for vertex--vertex interactions or \texttt{Op::VF} for vertex--face interactions, where the first letter refers to the first element of the pair and the second letter to the second.

\begin{cppcode}[]{}
    // Phase 2: define a pairwise interaction term
    problem.add_interaction_term<Op::VV>(
    [=] (auto pair_id, auto iter, auto var){
            // Access this pairs's mesh handles
            auto h0 = iter[0];
            auto h1 = iter[1];

            // Access the associated degrees of freedom as
            // active variables.
            auto x0 = var.active<ActiveT, 3>(h0);
            auto x1 = var.active<ActiveT, 3>(h1);

            // Simple distance-based interaction energy
            ActiveT dist = (x0 - x1).norm();
            ActiveT E = penalty_weight * dist;
            return E;});
\end{cppcode}

This programming model enables a unified treatment of mesh-local and interaction-based objectives within a single system. Users can freely combine elementwise terms with dynamically generated interaction terms of different types, while the system transparently handles derivative computation and sparse assembly. All operations, including sparse Jacobian and Hessian assembly, are performed entirely on the GPU, with no CPU--GPU memory round trips, as we discuss next.

\section{Design Principles}
\label{sec:design}

In this section, we describe the key design decisions that translate the user specification defined in our programming model (\S\ref{sec:prog_model}) into an efficient GPU implementation. These decisions are guided by the characteristics of mesh-based differentiation (e.g., small local degrees of freedom, sparse global coupling, and data-dependent interactions) and are aimed at achieving the design goals mentioned in \S\ref{sec:intro}, i.e., high performance while preserving generality and integration with existing GPU pipelines. Low-level implementation details are discussed in \S\ref{sec:details}.

\subsection{Forward-Mode Differentiation for Mesh Computations}
\label{sec:forward_mode}
A defining property of mesh-based objectives is that each local term $f_j(x_j)$ depends on a small and fixed number of degrees of freedom determined by the mesh element and its immediate neighborhood. As a result, the corresponding local derivatives (i.e., gradients $g_j \in \mathbb{R}^{k_j}$, Hessians $H_j \in \mathbb{R}^{k_j \times k_j}$, or Jacobians $J_j \in \mathbb{R}^{m_j \times k_j}$ for vector-valued terms) are small, dense objects whose sizes are independent of the global problem dimension. Global coupling arises only through sparse assembly across elements (Equation~\ref{eq:grad}).

This structure makes derivative computation \emph{local} and highly parallel. Each element can be differentiated independently and the only global operation is the accumulation of local contributions into sparse global structures. On the GPU, this accumulation can be implemented efficiently using parallel scatter or reduction primitives.

\paragraph{Forward-mode vs.\ reverse-mode AD}
Reverse-mode AD is asymptotically optimal for functions with many inputs and few outputs, a regime that dominates machine learning workloads and motivates its widespread adoption in modern AD frameworks~\cite{Martins:2021:EDO}. In that setting, dense connectivity across layers leads to high arithmetic intensity and amortizes the cost of constructing and traversing a global computation graph.

\begin{figure}
	\captionsetup[subfloat]{labelformat=empty}
	\subfloat[]{\fbox{\includegraphics[width=0.21\textwidth]{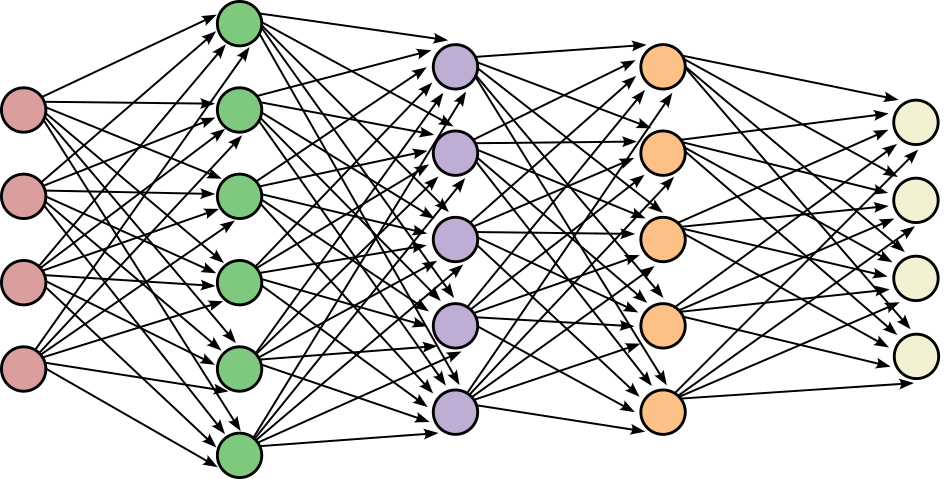}}}
	\subfloat[]{\fbox{\includegraphics[width=0.225\textwidth]{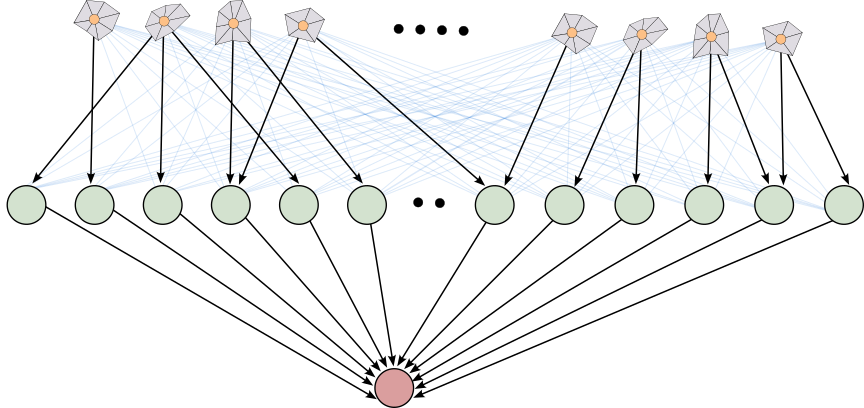}}}
	\caption{A typical network in machine learning workloads exhibits dense connectivity across many layers resulting in high arithmetic intensity and making reverse-mode AD a suitable choice (left). In contrast, mesh-based problems give rise to shallow, sparsely connected computation graphs (right) where each local function depends only on a small subset of inputs. This structure makes forward-mode AD more efficient for the GPU execution.}{}
	\label{fig:mesh_ad}
\end{figure}

Mesh-based differentiation operates in a different regime. As shown in Figure~\ref{fig:mesh_ad}, mesh objectives give rise to shallow, sparsely connected computation graphs in which each local function depends on only a small subset of inputs. Evaluating $f_j(x_j)$ and its derivatives involves relatively little computation but frequent memory access, making these workloads \emph{memory-bound}. In this context, the asymptotic advantage of reverse-mode AD does not translate into practical efficiency (see also Figure 4 in \citet{Schmidt:2022:TAD}).

Moreover, reverse-mode AD requires recording intermediate values during the forward pass and traversing a global computation graph in reverse. On GPUs, this typically entails increase in global memory traffic. While checkpointing~\cite{Griewank:2008:EDP} can reduce storage requirements, it introduces recomputation and synchronization overhead and does not eliminate dependence on global memory.

For these reasons, forward-mode AD is a better match for our setting. Since each local term has a small input dimension, forward-mode and reverse-mode have comparable theoretical cost at the element level. Forward-mode, however, enables derivatives to be computed \emph{on the fly} without constructing or traversing a global computation graph. This allows all intermediate quantities (i.e., primal values, directional derivatives, and local Jacobians or Hessians) to remain in registers or shared memory. Global memory accesses are limited to reading mesh data and writing the final sparse derivatives, aligning the differentiation process with the GPU memory hierarchy.

\subsection{Locality-aware Mesh Data Structures}
\label{sec:data_structure}
The effectiveness of forward-mode differentiation in the previous section relies on the assumption that the locality inherent in mesh-based objectives can be realized efficiently at runtime. While mesh energies are conceptually local, exploiting this locality on the GPU requires a data structure that treats locality as a first-class concern.

Traditional mesh data structures (e.g., halfedge-based~\cite{Mantyla:1988:ITS}, tuple-based~\cite{Brisson:1989:RGS} representations, or pointer-based adjacency lists) are designed primarily for flexibility and topological expressiveness on the CPU\@. In these structures, neighborhood traversal typically involves pointer chasing and irregular memory access, making it difficult to stage local computations efficiently or to guarantee coalesced access when traversed in parallel. As a result, these representations are a poor fit for GPU execution models that favor regular access patterns and on-chip reuse. Recent \emph{patch-based} mesh data structures (e.g., RXMesh~\cite{Mahmoud:2021:RAG} and MeshTaichi~\cite{Chang:2022:MAC}) address this gap by explicitly organizing mesh elements into small, independent patches that can be staged and processed cooperatively. For example, in RXMesh, mesh elements and their local neighborhoods are laid out to enable patch-local staging in shared memory, high cache reuse for accessing neighborhood element's attributes, and parallel traversal without pointer indirection. This organization aligns with the elementwise differentiation strategy described in \S\ref{sec:forward_mode} where local computations are dense but small, and global coupling arises only through sparse assembly.

We adopt RXMesh~\cite{Mahmoud:2021:RAG} as the underlying mesh representation and execution substrate for our system. Patch-local staging allows all data required for evaluating a local term (i.e., connectivity, attributes, and intermediate values) to be loaded once from global memory and reused across derivative computations. This organization enable coalesced access when iterating over neighborhoods and avoids pointer chasing, ensuring that parallel traversal remains efficient and scalable. Together, these properties make patch-based mesh data structures a natural foundation for locality-aware differentiation on the GPU\@.

\subsection{End-to-End GPU-Resident Execution}
\label{sec:gpu_resident}
The locality-aware data layout described above enables a second, equally important design principle, i.e., we execute all stages of evaluation and differentiation entirely on the GPU without CPU intervention or memory round trips. To achieve this, the system allocates only the data structures required by the problem specification and avoids materializing intermediate representations.

At initialization, we allocate buffers for the global objective, gradient, and sparse derivative matrices (Jacobian and Hessian) whose sizes and sparsity patterns are determined directly from the user's specification. We do not allocate additional intermediate storage, e.g., computation graphs, tapes, or dense derivative buffers. During execution, we generate all intermediate quantities required to evaluate local terms and their derivatives on the fly where they are confined to registers or shared memory.

This execution model applies uniformly across evaluation modes. For elementwise differentiation, RXMesh stores the minimal connectivity data needed for a patch while higher-level neighborhood queries (e.g., vertex one-rings) are generated dynamically within the kernel. Our local derivative computations similarly produce intermediate values only transiently, storing them  into registers or shared memory before contributing to the global sparse structures. This avoids unnecessary global memory traffic and eliminates the need to store intermediate per-element derivatives.

The same principle extends to matrix-free operations where we apply local derivative information directly to the input vector during kernel execution and we write only the final result to global memory. Likewise, passive evaluation modes reuse the same problem specification to compute objective values without derivative propagation, enabling efficient line search and related procedures without duplicating code or allocating additional buffers.

We handle changes to sparsity patterns (e.g., due to dynamically generated interaction terms) entirely on the GPU\@. As detailed in \S\ref{sec:details}, our system updates sparse matrix structures using a sequence of GPU kernels without synchronizing with the CPU or reallocating host-side data.

Finally, we store all assembled sparse matrices in standard Compressed Sparse Row (CSR) ensuring compatibility with high-performance GPU libraries, e.g., cuDSS and cuSolver. This design allows downstream linear solvers and optimization routines to operate directly on GPU-resident data, preserving end-to-end execution without CPU--GPU transfers.

\subsection{Decoupled Interaction Discovery and Evaluation}
\label{sec:interaction_design}
While \S\ref{sec:interaction_terms} introduces interaction terms from the user's perspective, we now describe the system-level design rationale behind their execution. Supporting dynamically generated interactions on the GPU requires addressing two distinct concerns: (i)~identifying which element pairs interact, and (ii)~evaluating the corresponding local terms and updating sparse derivative structures. We therefore adopt a two-stage design that explicitly decouples \emph{interaction discovery} from \emph{interaction evaluation}.

The first stage determines the set of nonzero couplings introduced by interactions and thus defines the additional sparsity that must be accommodated in the global Jacobian or Hessian. By separating discovery from evaluation, the system can detect changes to sparsity patterns early and allocate or update sparse data structures accordingly, without entangling this process with derivative computation. In the second stage, interaction terms are evaluated over the discovered pair set and simply populate these preallocated sparse structures with local derivative contributions. This design contrasts with alternative approaches that intertwine interaction detection with energy evaluation or rely on rebuilding global computation graphs when interactions change. 

\subsection{Atomic Assembly}
\label{sec:atomics}
Sparse differentiation in mesh-based problems requires two distinct forms of reduction. The first is a global reduction to evaluate the objective value $F(x)$ in Equation~\ref{eq:scalar_energy} which aggregates independent contributions. As this operation spans the full set of degrees of freedom, we use efficient GPU parallel reduction primitive, e.g., CUB~\cite{NVIDIA:2025:CUB}.

The second form arises during the assembly of global sparse derivatives. Each local term produces dense local contributions to the gradient, Jacobian, or Hessian, which must be scattered into global sparse structures. Because multiple local terms may contribute to the same global entry (particularly in the presence of mixed stencils and dynamically generated interactions) this assembly step requires synchronization across threads. We adopt \emph{atomic scatter-add operations} for sparse assembly. This choice is deliberate. Atomic updates provide a simple and correct mechanism for handling dynamic sparsity patterns without imposing additional constraints on the programming model. In practice, contention is typically low due to the locality of mesh interactions where each local term touches only a small number of global degrees of freedom and collisions are limited to shared mesh elements. Atomic accumulation is inherently nondeterministic and may lead to small run-to-run numerical differences, as is typical in GPU implementations. However, these differences usually do not affect qualitative behavior or convergence. Alternative strategies (e.g., graph coloring or sort-and-reduce schemes) either introduce a preprocessing overhead or restrict the class of admissible interaction patterns.

Atomics are used only where necessary. Local derivative computation for each term is performed privately within a thread or thread block, without synchronization. Atomic operations are applied only at the point of global accumulation. The same strategy is used for matrix-free operations (e.g., Hessian--vector products) where local contributions are computed independently and accumulated atomically into the output vector. This design balances generality and simplicity while maintaining high performance under dynamic sparsity.

\section{Implementation Details}
\label{sec:details}
This section describes the end-to-end execution pipeline of our system and explains implementation details whose realization is not immediately implied by the programming model or design principles.

\paragraph{Overall Pipeline.}

The system inputs a triangle surface mesh (which may be non-manifold and may contain disconnected components) together with a set of user-defined local terms specified using the programming model in \S\ref{sec:prog_model}. During initialization, we construct the RXMesh data structure and allocate all GPU-resident data required by the problem specification. This allocation includes (i)~user-declared mesh attributes and auxiliary fields and (ii)~global buffers for the objective value, gradient, and sparse derivative matrices (Jacobian and/or Hessian) whose structure is determined by the \texttt{Problem} configuration. The RXMesh data structures stores only the edge's vertices and the face's edges explicitly. All other query operations are performed on the fly per-patch in the shared memory. 

Execution proceeds by launching one GPU kernel per local term. Each kernel evaluates the user-defined computation independently over the relevant mesh elements, computes local derivatives using forward-mode AD, and accumulates the resulting contributions into global sparse structures (\S\ref{sec:sp_mat}, \S\ref{sec:local_derivative}). After all kernels finish, we compute the total objective value $F(x)$ in Equation~\ref{eq:scalar_energy} using a parallel reduction with CUB~\cite{NVIDIA:2025:CUB}.

The remainder of this section explains how we evaluate local derivatives, assemble sparse Jacobians and Hessians under dynamic sparsity, and implement these operations efficiently on the GPU\@.

\subsection{Sparse Hessian and Jacobian from Mesh Topology}
\label{sec:sp_mat}
A key property of mesh-based optimization problems is that the sparsity pattern of global derivatives is structured and predictable. In our formulation, optimization variables are defined per mesh element (e.g., vertices), each contributing $\texttt{VarDim}$ degrees of freedom. Let $n = \texttt{VarDim} \times |\mathbb{E}|$ denote the total number of degrees of freedom, where $\mathbb{E}$ is the set of optimization elements. The global Hessian $H \in \mathbb{R}^{n \times n}$ and has a block structure with $\texttt{VarDim} \times \texttt{VarDim}$ blocks.

\setlength{\columnsep}{7pt}
\begin{wrapfigure}[13]{r}{0.23\textwidth}
    \captionsetup[subfloat]{labelformat=empty}
    \vspace{-12pt}
    \fbox{\includegraphics[trim={2.5cm 2.5cm 2.0cm 2.5cm},clip, width=0.10\textwidth]{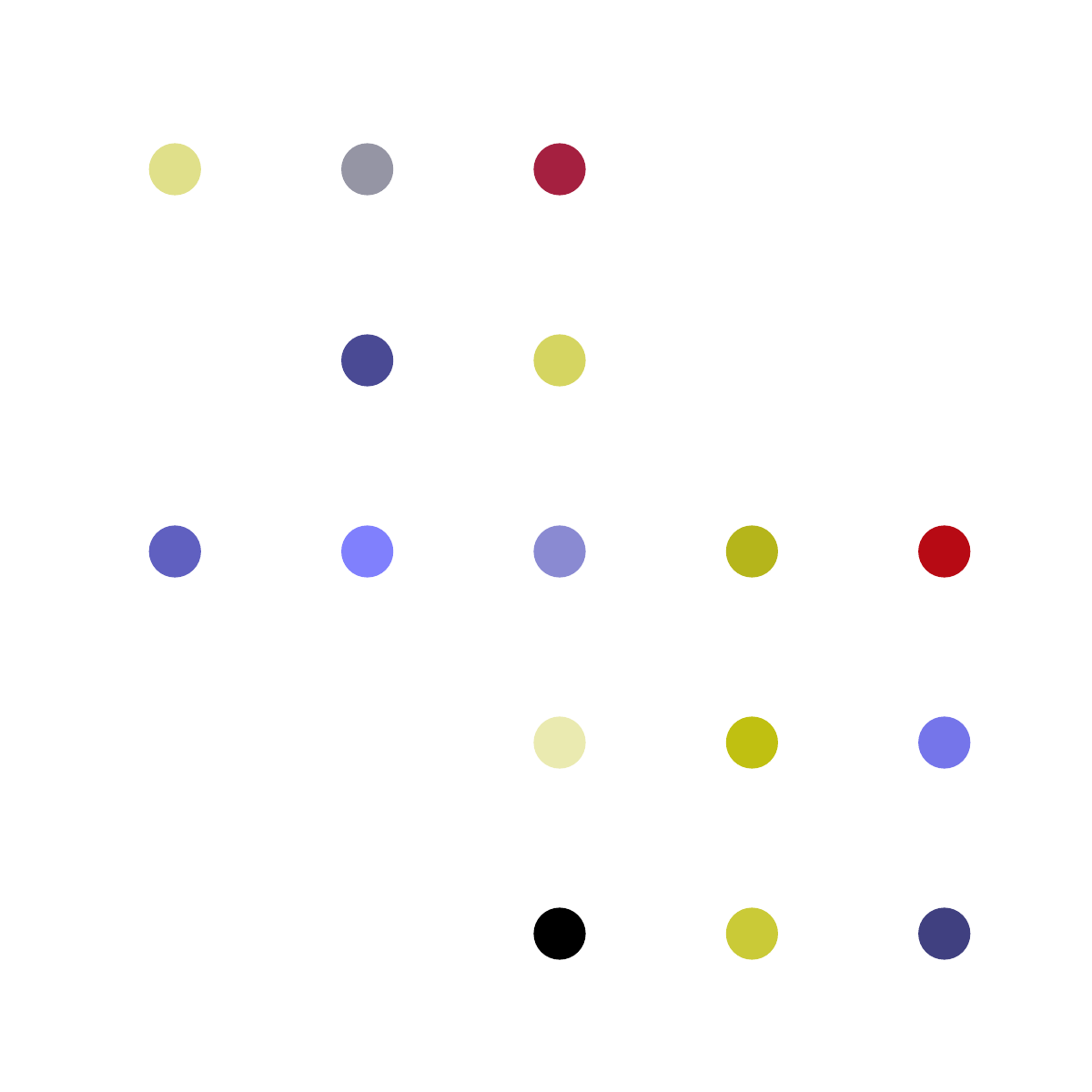}}
    \fbox{\includegraphics[trim={2.5cm 2.5cm 2.0cm 2.5cm},clip, width=0.10\textwidth]{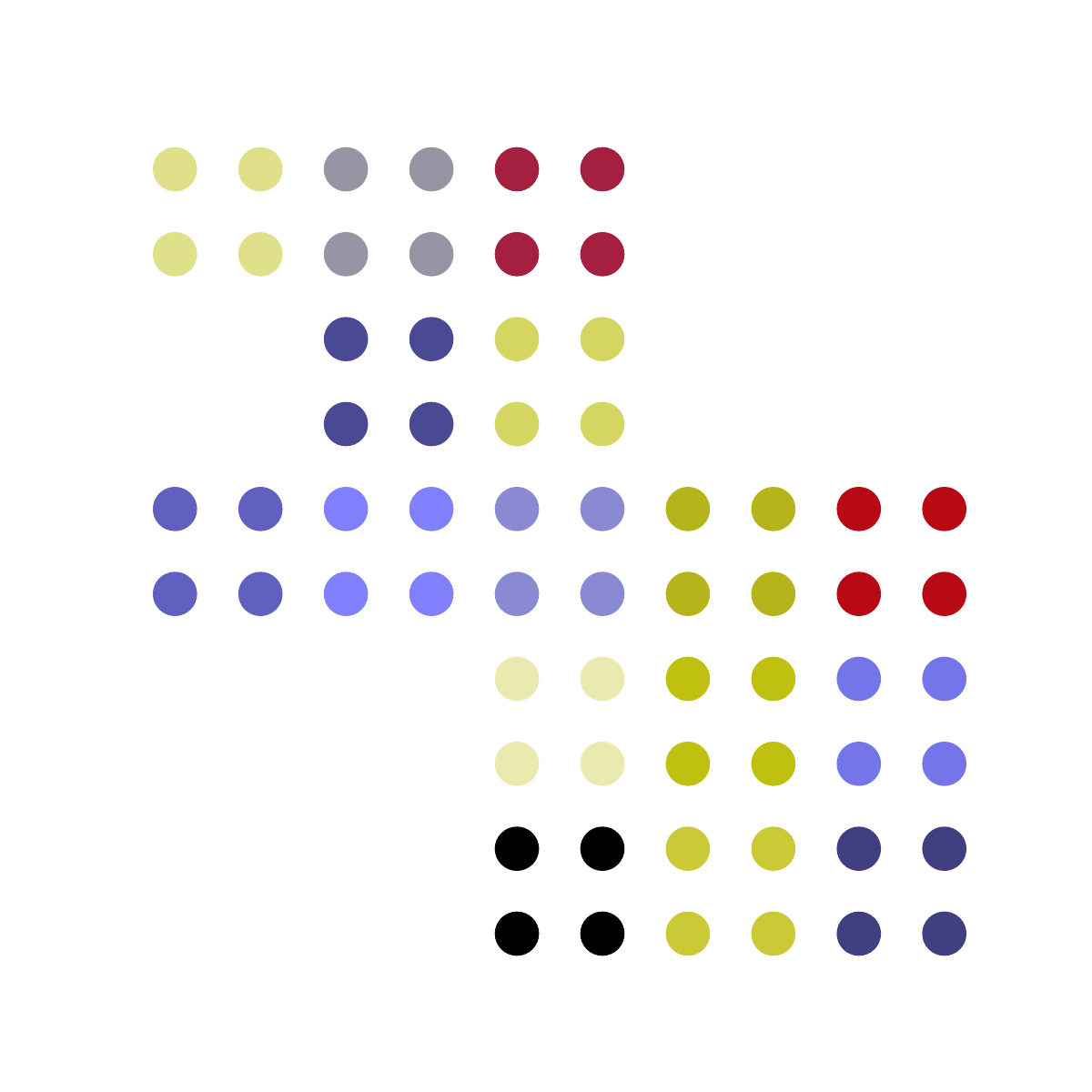}}
    \vspace{-10pt}
    \caption*{Sparsity pattern of the Hessian (right) can be derived from the mesh sparsity---graph Laplacian here (left)---where each entry of the mesh graph is replaced by a dense block of size $n\times n$ where $n$ is the dimension of the optimization variable. %
    }{}    
\end{wrapfigure}
The nonzero blocks of $H$ come directly from local energy terms. As described in \S\ref{sec:prog_model}, each term $f_j(x_j)$ depends only on a small subset of variables determined by the mesh stencil and contributes a dense local Hessian $H_j \in \mathbb{R}^{k_j \times k_j}$. When assembled according to Equation~\ref{eq:grad}, these local contributions produce nonzero blocks between pairs of optimization elements that participate in the same term. As a result, the block sparsity pattern of the global Hessian mirrors the adjacency structure induced by the mesh connectivity and the declared neighborhood queries (see inset).

Rather than discovering this sparsity pattern dynamically, we exploit its predictability to preallocate the Hessian structure before differentiation. We store the global Hessian in a Compressed Sparse Row (CSR) format. During initialization, we analyze the mesh topology together with the neighborhood access patterns declared in the user-defined \texttt{Problem} (\S\ref{sec:prog_model}) to enumerate all potential element--element interactions. This analysis runs in parallel on the GPU and is followed by a prefix-sum scan to compute global offsets and total storage requirements. When the user calls \texttt{problem.eval\_terms()}, each kernel computes dense local Hessian blocks and accumulates them into the preallocated global structure using atomic additions. This way, we avoid dynamic memory allocation during differentiation, reduce runtime overhead, and ensure stable GPU performance.

\paragraph{Sparse Jacobians.}
For vector-valued objectives $F:\mathbb{R}^n \rightarrow \mathbb{R}^m$, the global Jacobian $J \in \mathbb{R}^{m \times n}$ follows an analogous structure (Equation~\ref{eq:jac}). Each local term $f_j$ produces a dense local Jacobian $J_j \in \mathbb{R}^{m_j \times k_j}$ with respect to its local degrees of freedom. Using the selection matrix $S_j$, each local Jacobian is expanded to $J_j S_j \in \mathbb{R}^{m_j \times n}$, coupling a small subset of variables to a contiguous segment of the output vector.

In our implementation, we assign each term ownership of a disjoint block of rows via the mapping $P_j$ and assemble the global Jacobian by stacking these expanded local contributions along the row dimension. As in the Hessian case, the column sparsity of $J$ is determined solely by mesh connectivity and the declared stencil. We therefore preallocate the Jacobian in CSR format and, during evaluation, each kernel writes its local Jacobian blocks directly into the global matrix without materializing dense intermediates or performing CPU-side assembly.

\subsection{Local Derivative Evaluation}
\label{sec:local_derivative}
Our system evaluates derivatives by executing each user-defined term independently over the mesh using forward-mode automatic differentiation. For each term, we first compute (or simply read) the required connectivity information in shared memory using the RXMesh pipeline, e.g., computing a face's three vertices. The user specifies this required connectivity information in the definition of their term (Line~\ref{line:op} in Listing~\ref{list:api}). Given this connectivity information, we then execute the user-defined term by launching one thread per mesh element (e.g., one thread per edge in Listing~\ref{list:api}) and passing the mesh element handle and an iterator over the required connectivity. In the case of unary operations (i.e., \texttt{Op::V}, \texttt{Op::E}, \texttt{Op::F}), we pass only a mesh element handle.

At this point, the objective (and all auxiliary variables) reside in global memory as plain floating-point values. When the user fetches the objective as in Line~\ref{line:lift} in Listing~\ref{list:api}, we load the objective from global memory into registers and lift the variables into our custom \texttt{ActiveT} type which wraps scalar values and tracks derivatives using \emph{compact local indexing}. Internally, an \texttt{ActiveT} instance stores the primal value and its derivatives with respect to a small, fixed set of local variables determined at compile time by the stencil size. We overload arithmetic operations for \texttt{ActiveT} to propagate values and derivatives via the chain rule, producing dense local gradients and Hessians without constructing an explicit computation graph. The indexing depends on the stencil type which the kernel knows at compile time, allowing the compiler to keep derivative storage in registers. For example, evaluating the edge energy in Listing~\ref{list:api} lifts the two vertex positions into \texttt{ActiveT}, propagates derivatives alongside values, and produces a local gradient $g \in \mathbb{R}^{k}$ and Hessian $H \in \mathbb{R}^{k \times k}$, where $k = 2 \times VarDim$. Knowing the local derivative sizes at compile time lets the compiler optimize derivative evaluation without requiring a highly specialized kernel. The same logic applies to pairwise interaction terms where the stencil size is 2.

The return value of the user lambda is another \texttt{ActiveT} value that stores the derivatives of the element contribution (the edge in Listing~\ref{list:api}) with respect to the local degrees of freedom of the stencil members (the vertices in Listing~\ref{list:api}). For unary terms, each thread owns a single mesh element. In this case, the kernel writes the local energy contribution and accumulates the gradient and diagonal Hessian block directly, without synchronization. For neighborhood-based terms (e.g., \texttt{Op::FV}, \texttt{Op::EV}), each thread evaluates a dense local derivative block over the stencil and scatters these contributions into the global gradient and sparse Hessian using atomic additions. This separation ensures that atomics are used only at the point of global accumulation while all local differentiation remains private to the thread.

The same mechanism supports multiple execution modes controlled by \texttt{ActiveT}. In gradient-only mode, the kernel instantiates an active scalar that tracks first derivatives only. In passive mode, \texttt{ActiveT} yields plain floating-point values with no derivative tracking. In matrix-free mode, after computing the local Hessian, we apply it directly to an input vector and accumulate the result into the output vector without materializing the global Hessian. Across all modes, our system confines differentiation to local, fixed-size computations and produces global derivatives solely through sparse accumulation. %

\subsection{Dynamic Updates of Sparse Structures}
\label{sec:dynamic_sparsity}
Consistent with our end-to-end GPU-resident execution principle (\S\ref{sec:design}), we perform all updates to the derivative structures fully in parallel on the GPU without staging data on the CPU\@. Although CSR is not an ideal format for incremental insertion, we deliberately adopt it to satisfy our integration goal, i.e., maintaining compatibility with high-performance GPU sparse solvers such as cuDSS and cuSolver\@. This choice allows our system to interface directly with optimized solver backends while still supporting dynamic changes in sparsity.

When a user introduces new interaction pairs during execution (e.g., via collision detection as shown in \S\ref{sec:interaction_terms}), the user inserts only pairs of mesh handles, while the system internally expands these into the corresponding global row and column indices for the dense $VarDim \times VarDim$ blocks contributed by each interaction. This expansion depends on the element indices and variable dimensionality allowing the resulting sparse updates to be expressed as a set of new nonzero entries in Coordinate List (COO) form.

Algorithm~\ref{alg:csr_insert} in Appendix~\ref{appendix:csr_update} describes how we incorporate these new entries into the global Hessian structure entirely on the GPU\@. We maintain two CSR matrices in a double-buffered configuration and preallocate sufficient storage to accommodate both the initial sparsity pattern and a bounded number of future interactions. Upon insertion, we compute updated row offsets, copy existing entries, and insert the new nonzeros in parallel, after which the active and inactive buffers are swapped. Our design avoids dynamic memory allocation, preserves solver compatibility, and ensures that sparsity updates remain scalable and fully GPU-resident.

\section{Applications}
\label{sec:apps}
We evaluate our system on seven different applications in mesh-based simulation, geometry processing, and optimization, each highlighting a different capability or performance aspect. All experiments are conducted on an NVIDIA RTX 4090 GPU with 24~GB memory, using CUDA 13 and Visual Studio 2022 on a system with an Intel\textsuperscript{\textregistered} Core\texttrademark~i9-14900KF (24-core, 4.4~GHz) and 64~GB RAM. Reported timings exclude preprocessing and data structure initialization.%

In summary, the mass-spring cloth simulation application (\S\ref{sec:spring}) evaluates gradient and sparse Hessian computation for second-order methods with multiple energy terms, compared against an optimized PyTorch implementation. Mesh parameterization (\S\ref{sec:param}) tests our system's ability to compute efficient Hessian-vector products in scenarios where explicit Hessian construction is unnecessary. Manifold optimization (\S\ref{sec:manifold}) highlights the system's flexibility in more complex tasks and benchmarks first-order derivative performance in the L-BFGS algorithm. Polyvector design (\S\ref{sec:polyvector}) evaluates our system's performance for vector-valued problems. ARAP mesh deformation (\S\ref{sec:arap}) compares our system against Thallo~\cite{Mara:2021:TSF} for non-linear least square problems. The smoothing application (\S\ref{sec:area}) serves as a first-order optimization benchmark, comparing our gradient computation to various baselines and to manual derivative implementation. Finally, the elastic simulation example (\S\ref{sec:elastic_sim}) explores our system's performance for dynamically changing Hessian sparsity. 

\subsection{Cloth Simulation with Mass Spring Model}
\label{sec:spring}
\begin{figure}
    \centering
    \includegraphics[width=0.48\textwidth]{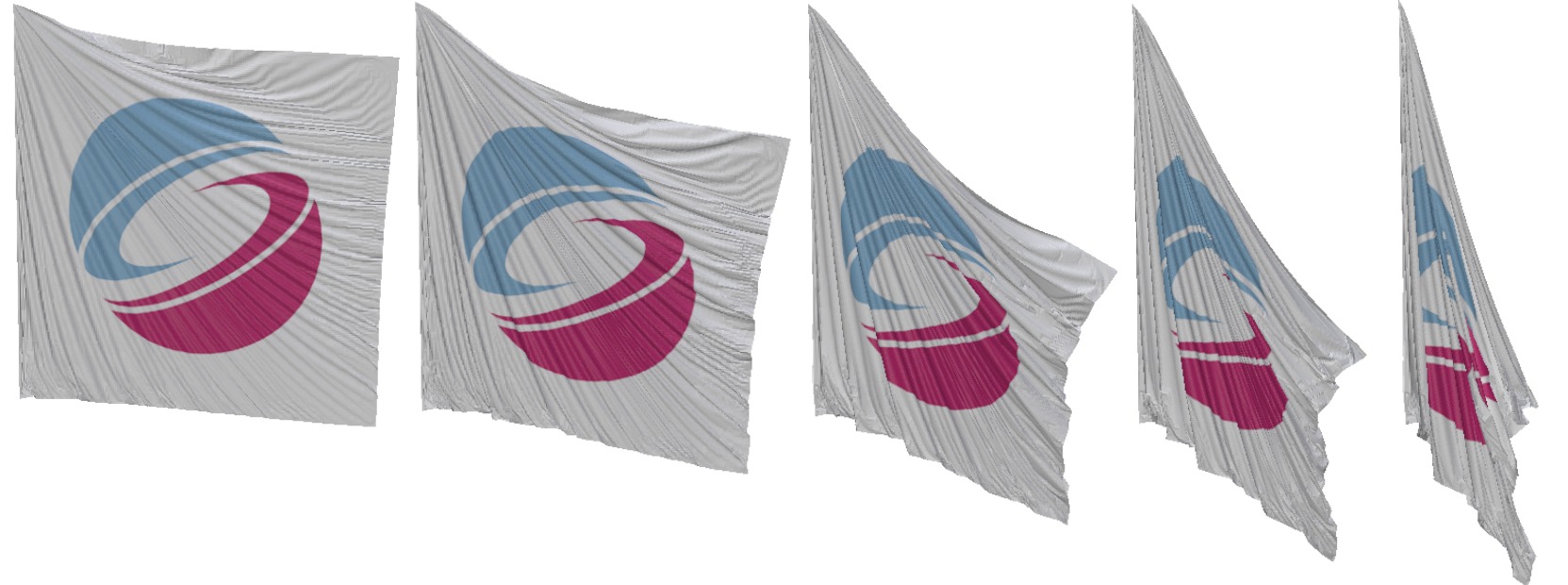}
    \caption{Cloth simulation driven by a mass--spring system with inertial energy, elastic spring potentials, and gravity. The simulation relies on sparse Hessian computation of these energies at each time step.}
    \label{fig:ms_flag}
\end{figure}

Our first application is the classical mass-spring system for cloth simulation where mesh edges act as springs and the system evolves under inertia and external forces~\cite{Provot:1995:DCI} (see Figure~\ref{fig:ms_flag}). Each timestep minimizes a composite energy consisting of inertial, elastic potential, and gravitational terms. Using implicit Euler integration~\cite{Li:2024:PBS}, we optimize the total energy:
$$
    E(x) = \frac{1}{2} \left\| x - \left(x^n + h v^n \right) \right\|^2_M + h^2 P(x),
$$
where $x$ is the a vector of the next vertex position, $x^n, v^n$ are current positions and velocities, $M$ is the lumped mass matrix, $h$ is the timestep, and $P(x)$ is a potential energy function.

The potential $P(x)$ includes:
\begin{itemize}
    \item Spring energy: For edge $e = (i,j)$,
          $$
              P_e(x) = l^2 \frac{1}{2}k \left( \frac{ \| x_i - x_j \|^2 }{l^2} - 1 \right)^2,
          $$
          where $x_i, x_j$ are the positions of the edge endpoints, $l$ is the rest length, and $k$ is a spring stiffness coefficient.
    \item Gravitational energy: $P_g(x) = -x^\top M g$, with $g$ the constant gravity vector.
\end{itemize}

To optimize $E(x)$, we use Newton's method with backtracking line search. Users define per-edge and per-vertex energies. Then, our system computes gradients and sparse Hessians automatically. We then use the resulting gradient and Hessian to solve the linearized system (using cuDSS~\cite{NVIDIA:2025:cudss}) and update the solution. To improve robustness under large deformations, we apply \emph{Hessian filtering}~\cite{Teran:2005:RQF} by clamping negative eigenvalues of individual Hessian energy terms to a small positive value before assembling the matrix, ensuring descent directions and stable convergence.

We compare our system to two alternative cloth simulation implementations,  summarized in Table~\ref{tab:spring} and detailed below.

\begin{table}
    \setlength{\tabcolsep}{5.0pt}
    \caption{Mass-spring performance. We report only the time per time step spent evaluating gradients and Hessians, excluding other components such as the linear solver. Results are averaged over 1000 time steps on meshes of varying sizes, identified by their number of vertices. \label{tab:spring}}
    \begin{small}
        \begin{center}
            \begin{tabular}{cccc}
                \toprule
                \# V     & PyTorch  (ms) & \texttt{IndexedSum} (ms) & Ours (ms)     \\
                \midrule
                $10^2$   & 269.8         & 17.07                    & \textbf{0.18} \\ %
                $50^2$   & 7,205.6       & 18.72                    & \textbf{0.22} \\ %
                $100^2$  & 29,516.76     & 11.74                    & \textbf{0.25} \\ %
                $500^2$  & OOM           & 15.73                    & \textbf{3.09} \\ %
                $1000^2$ & OOM           & 39.67                    & \textbf{11.7} \\ %

                \bottomrule
            \end{tabular}
        \end{center}
    \end{small}
\end{table}

\paragraph{PyTorch:}
In PyTorch, users define local energies as Python functions and compose the total energy via vectorized summation. We used (the reverse-mode) \texttt{autograd} and \texttt{torch.func.hessian} to compute the gradients and Hessians, respectively. While this supports AD out of the box similar to our system, it produces \emph{dense} Hessians, ignoring problem sparsity. By exploiting sparsity and GPU locality, our system achieves over two orders of magnitude speedup. As a consequence of PyTorch returning dense Hessians, Newton's method relies now on dense Cholesky factorization, incurring significant memory overhead and poor scalability.

\paragraph{IndexedSum:}
To improve upon the naive PyTorch baseline, we compare against \texttt{IndexedSum}---a recent PyTorch-based library for sparse Hessian computation.\footnote{\url{https://github.com/alecjacobson/indexed_sum}} \texttt{IndexedSum} provides a lightweight wrapper around PyTorch \texttt{autograd} that constructs sparse Hessians from elementwise contributions using a gather-compute-scatter model. This approach closely mirrors our system, i.e., the user defines a local energy per element and the framework automatically assembles the global energy, gradient, and sparse Hessian. While our work relies on forward-mode AD, \texttt{IndexedSum} performs a vectorized dense AD on the local Hessians using reverse-mode AD.

While \texttt{IndexedSum} reduces memory overhead and is significantly faster than naive PyTorch, it still performs local derivative computations through global memory accesses. In contrast, our system uses forward-mode AD to keep all local computation in registers and shared memory, issuing only a single global read (mesh data) and a single global write (gradients and Hessians). As illustrated in Figure~\ref{fig:ms_breakdown}, this design reduces differentiation cost enough to shift the bottleneck to the linear solver and leads to a \textbf{5.1$\times$ speedup} over \texttt{IndexedSum} on meshes with approximately $10^6$ vertices.

\setlength{\columnsep}{7pt}
\begin{wrapfigure}[12]{r}{0.23\textwidth}
    \captionsetup[subfloat]{labelformat=empty}
    \vspace{-12pt}    
    \includegraphics[width=0.24\textwidth]{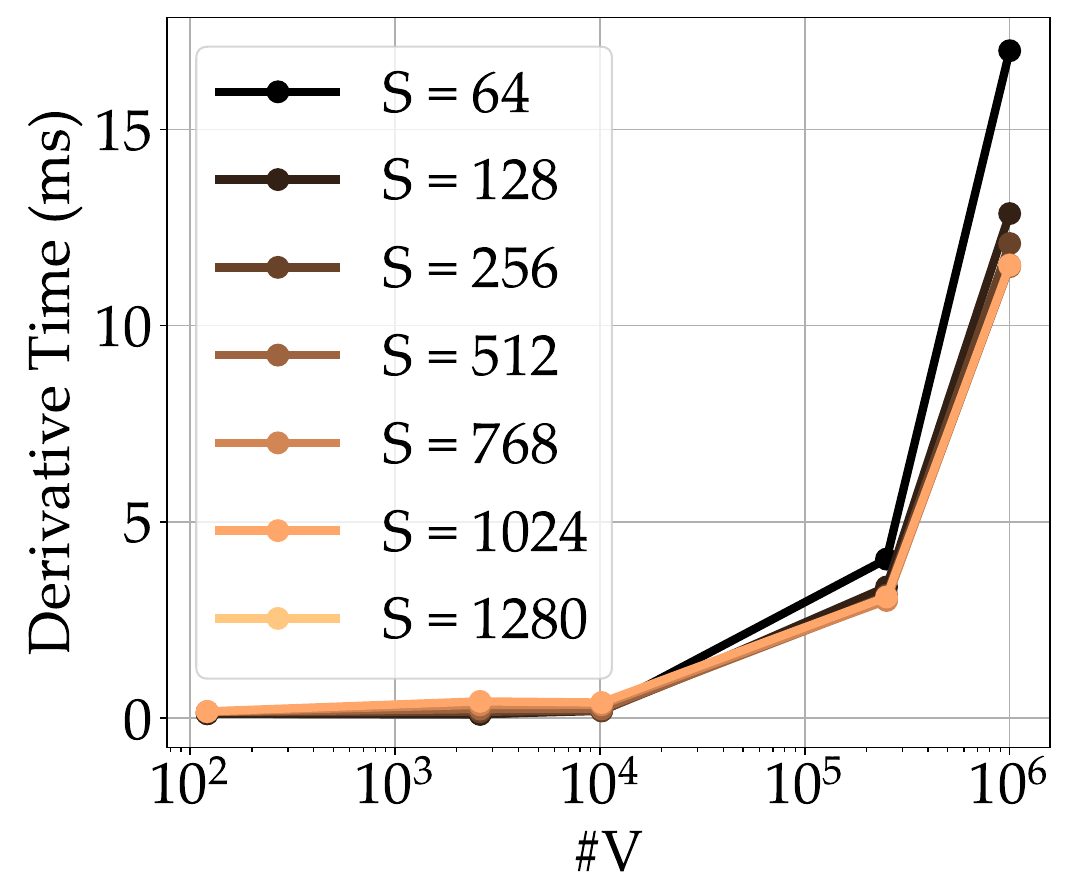}
    \vspace{-16pt}
    \caption*{Impact of patch size on differentiation time}{}
    \label{fig:patch_size_study}   
\end{wrapfigure}

We also study the overhead of patch construction. Generally, patch construction time is negligible and amounts to approximately 50 ms for a mesh with one million vertices. This preprocessing cost is incurred once at initialization and is amortized across all subsequent iterations of the simulation or optimization. We also study performance as a function of patch size, defined as the target number of faces provided to the $k$-means algorithm. The inset shows the differentiation time, including derivative computation and accumulation into the global Hessian, for different meshes and different patch sizes. We observe that beyond a moderate patch size (around 256 faces), performance becomes largely insensitive to this parameter. Based on this result, we choose a target patch size of 512 as the default for all applications.

\subsection{Mesh Parameterization with Symmetric Dirichlet Energy}
\label{sec:param}

\begin{figure}
    \centering
    \includegraphics[width=0.48\textwidth]{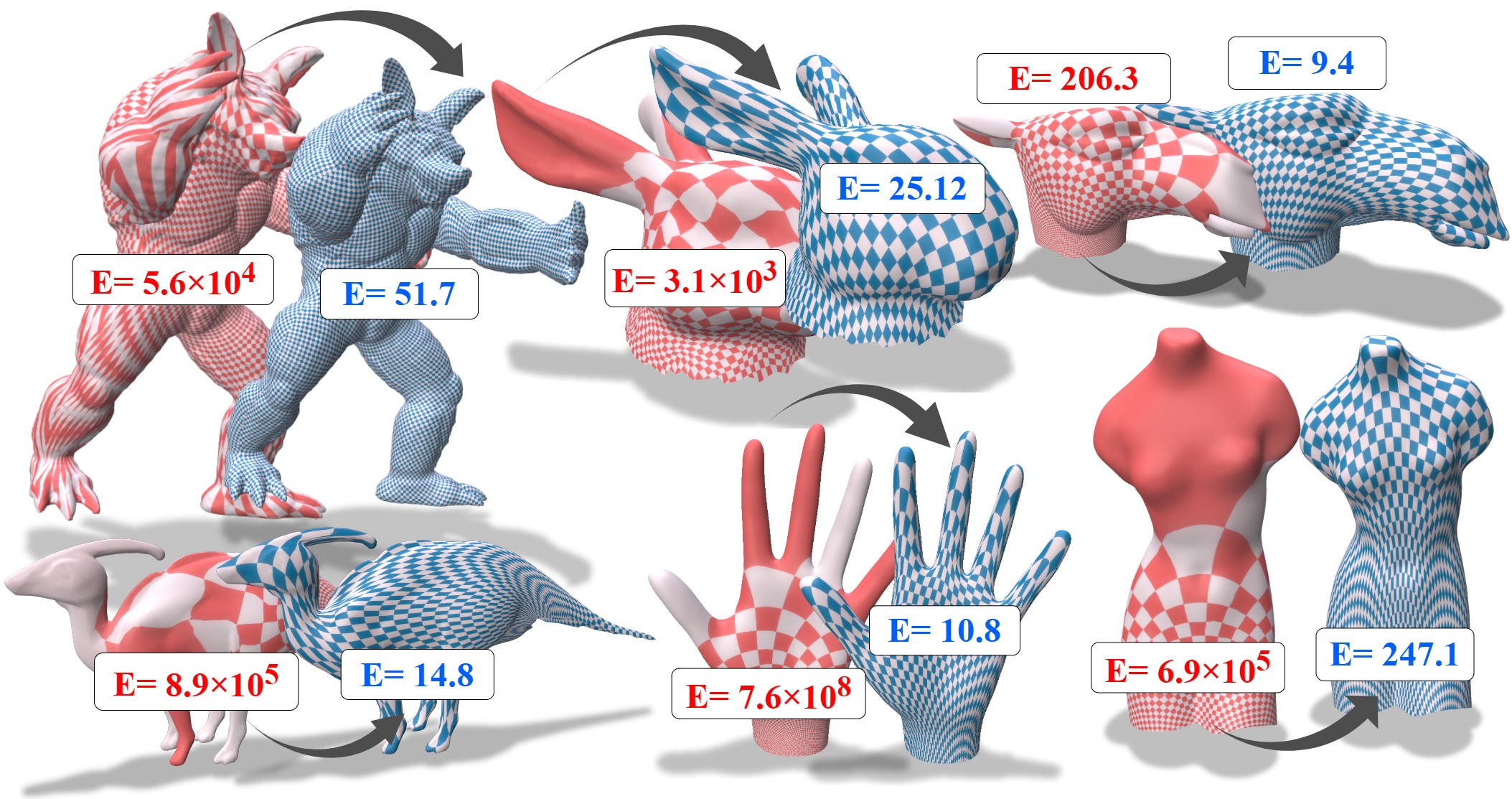}
    \caption{Parameterization results for several meshes using symmetric Dirichlet energy minimization. Red shows the initial parameterization and blue the optimized result. The reported values $E$ denote the total symmetric Dirichlet energy before and after optimization. Our system computes gradients and Hessian-vector products locally on the GPU and solves the problem using Newton's method with line search.}
    \label{fig:param_fig}
\end{figure}

This application minimizes face-based distortion in mesh parameterization. The goal is to assign each 3D vertex a 2D coordinate (UV map) such that the mapping is as distortion-free as possible---critical for tasks like texture mapping, remeshing, and geometric modeling. We use the (area-scaled) symmetric Dirichlet energy~\cite{Smith:2015:BPW, Schreiner:2004:ISM}, which penalizes both stretching and shrinking of each triangle uniformly. Given a triangle mesh $\mathcal{M} = (\mathcal{V}, \mathcal{T})$ and an initial 2D map $x \in \mathbb{R}^{2|\mathcal{V}|}$, the objective is:
$$
    f(x) = \sum_{t \in \mathcal{T}} \text{area}_t\cdot \left( \left\| J_t(x) \right\|_{F}^2 + \left\| J_t(x)^{-1} \right\|_{F}^2 \right),
$$
where $\| \cdot \|_{F}$ is the Frobenius norm, $J_t(x)$ is the $2 \times 2$ Jacobian of the affine UV mapping per triangle, computed as
$
    J_t = [b - a \quad c - a] \cdot [b_r - a_r \quad c_r - a_r]^{-1},
$
with $ a, b, c $ as UV coordinates and $ a_r, b_r, c_r $ the corresponding rest-pose 3D positions expressed in the triangle local 2D coordinates system.

We implement this energy by defining a per-triangle energy term using our system. The user provides the local energy expression and our system computes the global energy, gradient, and (optionally) the Hessian. We use Newton's method with line search for the minimization (see Figure~\ref{fig:param_fig}). Unlike the previous example, we avoid explicit Hessian assembly and adopt a matrix-free strategy. We use conjugate gradient (CG) which requires only Hessian-vector products. These are computed in our system by deriving localized per-element contributions without forming or storing the global Hessian.

For performance benchmark, we compare against a PyTorch implementation that computes Hessian-vector products using the \emph{double backward} mechanism via \texttt{torch.autograd.grad}. This approach evaluates the directional derivative of the gradient by applying a second backward pass without explicitly forming the Hessian. Both implementations yield matching energy trajectories. Across various mesh resolutions, our system achieves a geometric mean \textbf{2.29$\mathbf{\times}$ speedup} over PyTorch (see Table~\ref{tab:slim}). This gain results from our design which avoids dynamic graph construction and reduces global memory traffic through localized computation.
\begin{table}
    \setlength{\tabcolsep}{5.0pt}
    \caption{Performance of the Newton solver using matrix-free Conjugate Gradient (CG) for mesh parameterization. Instead of constructing the full Hessian, we compute Hessian-vector products on-the-fly during each CG iteration. The reported time is the average per CG iteration. \label{tab:slim}.}
    \begin{small}
        \begin{center}
            \begin{tabular}{cccc}
                \toprule
                \# V ($\times 10^6$)  & PyTorch (ms) & Ours (ms) & Speedup               \\
                \midrule
                1.55 $\times 10^{-3}$ & 0.039        & 0.026     & \textbf{1.4}$\times$  \\
                0.16                  & 0.27         & 0.108     & \textbf{2.4}$\times$  \\
                0.56                  & 0.896        & 0.383     & \textbf{2.3}$\times$  \\
                1.52                  & 2.475        & 0.817     & \textbf{3.03}$\times$ \\
                1.83                  & 2.987        & 1.102     & \textbf{2.71}$\times$ \\
                \bottomrule
            \end{tabular}
        \end{center}
    \end{small}
\end{table}

\subsection{ARAP Mesh Deformation}
\label{sec:arap}

This application evaluates our system on a classical non-linear least-squares problem in geometry processing, i.e., as-rigid-as-possible (ARAP) mesh deformation. ARAP is widely used for interactive shape editing, where users prescribe positional constraints on a subset of vertices and the system computes a globally consistent deformation that preserves local rigidity. From a system perspective, ARAP is representative of vector-valued, non-linear least-squares problems with coupled unknowns, resulting into a sparse Jacobian.

We formulate ARAP using the embedded deformation model. Unlike traditional ARAP implementations that alternate between local rotation projection and global solves~\cite{Sorkine:2007:ARAP}, we adopt a joint optimization over translations and rotations. This formulation is chosen specifically to enable a direct comparison against Thallo~\cite{Mara:2021:TSF}. Note that we do not advocate this parameterization as the preferred ARAP formulation in general.

In this formulation, each vertex $i$ is associated with a translation $o_i \in \mathbb{R}^3$ and a local rotation $R_i \in \mathbb{R}^{3 \times 3}$. Given rest positions $u_i$, the deformed positions are $x_i = u_i + o_i$. The total energy is
\begin{equation}
    E = w_{\text{fit}}E_{\text{fit}} + w_{reg}E_{\text{reg}} + w_{\text{rot}}E_{\text{rot}}.
\end{equation}
where $w_{\text{fit}}, w_{reg},$ and $w_{\text{rot}}$ are weights. The fitting term enforces user constraints as:
\begin{equation}
    E_{\text{fit}} =
    \sum_{i \in V}
    \begin{cases}
        \|o_i\|^2,       & i \text{ fixed},     \\
        \|o_i - c_i\|^2, & i \text{ displaced}, \\
        0,               & \text{otherwise},
    \end{cases}
\end{equation}
where $c_i$ denotes a user-specified displacement. The regularization term enforces local rigidity by matching deformed edge vectors to rotated rest-pose edges:
\begin{equation}
    E_{\text{reg}} =
    \sum_{(i,j)\in E}
    \left\|
    (x_j - x_i) - R_i (u_j - u_i)
    \right\|^2.
\end{equation}
Finally, the rotation term softly enforces orthonormality of the rotation matrices using column dot products and norms:
\begin{equation}
    \begin{aligned}
        E_{\text{rot}} =
        \sum_{i \in V}
        \Big(
         & \, c_{0,i}^\top c_{1,i}
        + c_{0,i}^\top c_{2,i}
        + c_{1,i}^\top c_{2,i}     \\
         & + (\|c_{0,i}\|^2 - 1)
        + (\|c_{1,i}\|^2 - 1)
        + (\|c_{2,i}\|^2 - 1)
        \Big)^2 .
    \end{aligned}
\end{equation}
where $c_{k,i}$ denotes the $k$-th column of $R_i$.

We minimize this energy using Gauss--Newton, solving a sparse normal equation at each iteration. Users specify per-vertex and per-edge residuals, and our system automatically constructs the global residual vector, Jacobian, and sparse normal matrix. %

In this application, we compare against Thallo which is a domain-specific system for nonlinear least-squares problems that relies on auto-scheduling to select an efficient strategy for forming and applying $J^\top J$, including whether to materialize the Jacobian or intermediate products. Similar to our system, Thallo allows users to define energies and then automatically generates sparse Jacobians. Both our system and Thallo use preconditioned conjugate gradient to solve the linear system.

\begin{figure}
    \centering
    \includegraphics[width=0.48\textwidth]{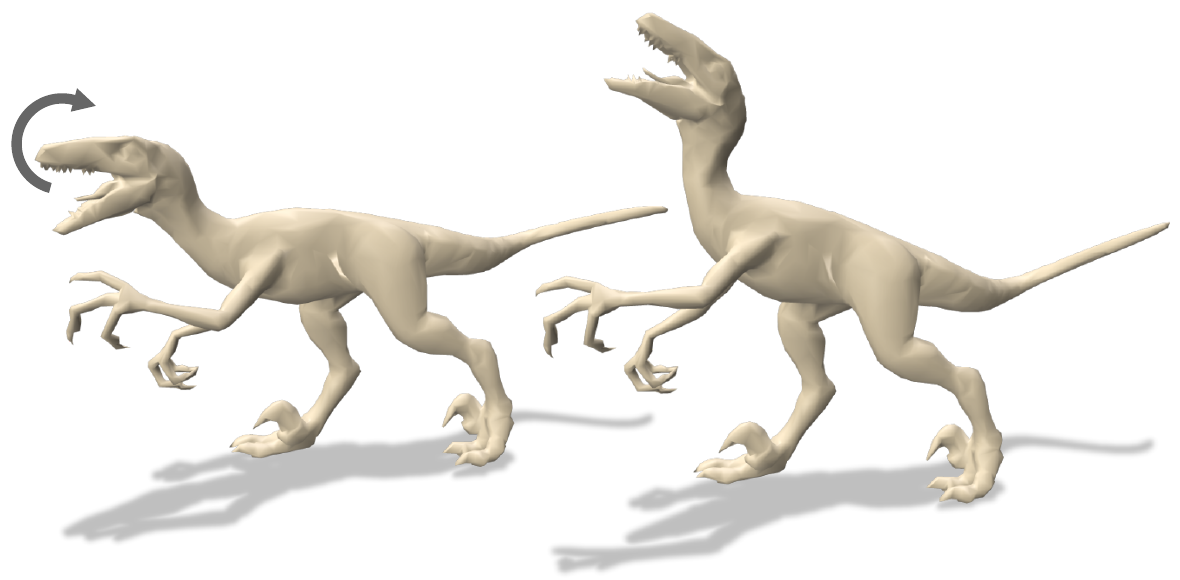}
    \caption{ARAP deformation solved via nonlinear least squares optimization. The input mesh (left) is deformed by minimizing a collection of local energy terms using Gauss--Newton (right).}
    \label{fig:arap_fig}
\end{figure}

For this application, Thallo's auto-scheduler selects a strategy that avoids materializing $J^\top J$. Despite this, our optimized implementation achieves higher performance, resulting in a $2.01\times$ speedup over Thallo (Figure~\ref{fig:arap_fig}). On average, Thallo requires 9.998 ms per Gauss--Newton iteration, whereas our solver requires 4.979 ms per iteration.

\subsection{Integrable Polyvector Fields}
\label{sec:polyvector}
\begin{figure}
    \centering
    \includegraphics[width=0.48\textwidth]{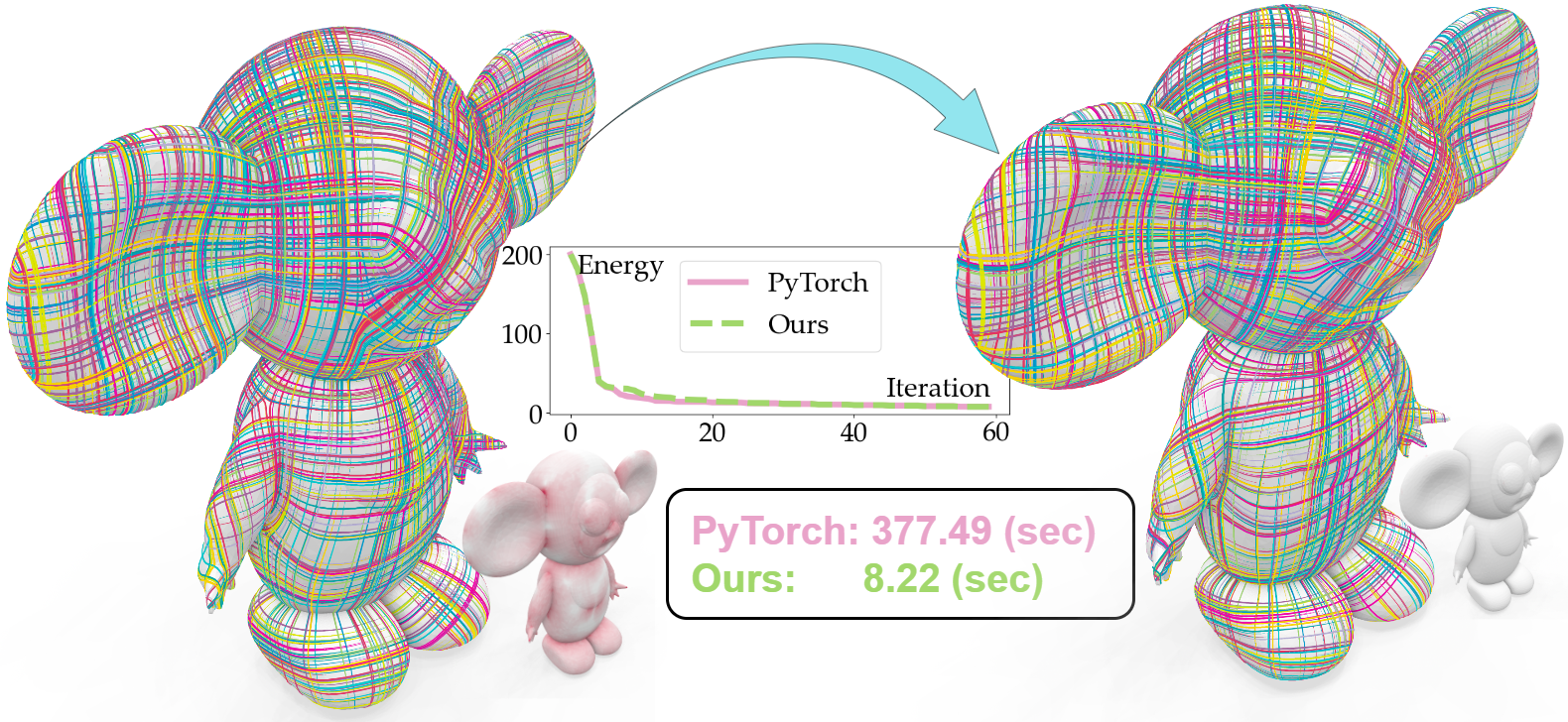}
    \caption{We project an input frame field (left) onto the space of curl-free, integrable fields, producing a field that induces an inversion-free surface parameterization (right)~\cite{Diamanti:2015:IPF}. The plot shows the energy decay over 60 iterations for PyTorch and our system which converge to the same solution. The inset visualizes the PolyCurl term before and after optimization. Our system completes the full solve in \textbf{8.22 seconds}, compared to \textbf{377.49 seconds} for PyTorch, highlighting the performance impact of explicitly materializing and reusing sparse Jacobians.}
    \label{fig:polyvector}
\end{figure}

Similar to the previous application, we use integrable frame field optimization as another nonlinear least-squares application that exercises several aspects of our system. Following \citet{Diamanti:2015:IPF}, the goal is to take an input frame field defined on a triangle mesh and project it onto the space of curl-free fields, which are locally integrable and therefore correspond to gradients of valid surface parameterizations. Each face $f$ carries two tangent vectors $(\alpha_f, \beta_f)$ represented as complex numbers in the local tangent basis. To enforce curl-freeness, smoothness, and order preservation, we use a collection of per-edge and per-face energy terms, resulting in a nonlinear objective of the form $E(x) = \sum_j r_j(x)^2$, where $x \in \mathbb{R}^{4|\mathcal{F}|}$ stacks the real and imaginary parts of the two vectors per face (see Equation~16 in \citet{Diamanti:2015:IPF}).
We solve this problem using Gauss--Newton (GN) iterations, which require evaluating the residual vector and assembling the sparse Jacobian $J = \partial r / \partial x$ at every iteration. For details of the per-face and per-edge energy definitions, we refer the reader to \citet{Diamanti:2015:IPF}.

This application highlights a few capabilities of our system. The unknowns are defined per face rather than per vertex, exercising native support for face-centered variables. The formulation also operates directly in complex arithmetic, closely mirroring the original mathematical description of the method and demonstrating that our system can express and differentiate energies involving complex-valued quantities without special handling. In our implementation, we directly use the \texttt{thrust::complex} type from the Thrust library~\cite{NVIDIA:2025:thrust}. While several energy terms are defined per face, most coupling terms are evaluated per edge and scatter contributions to adjacent faces through edge-to-face adjacency. This access pattern maps to our \texttt{Op::EF} operator which we use for parallel assembly of both residuals and Jacobian entries.

At iteration $k$ of the GN method, we solve the linear system $J(x_k)^{\mathsf{T}} J(x_k)\,\Delta_k = - J(x_k)^{\mathsf{T}} r(x_k)$ using a conjugate gradient solver, and update the solution as $x_{k+1} = x_k + \Delta_k$. We set the maximum number of iterations in conjugate gradient to 500. We compare the performance of our system on this application against PyTorch. In PyTorch, and similarly in other ML-oriented systems, the recommended approach is to avoid explicitly materializing the Jacobian and instead rely on Jacobian--vector products (JVPs) and vector--Jacobian products (VJPs). This design choice is primarily due to the difficulty of constructing sparse Jacobians efficiently in these frameworks. Accordingly, we implement the conjugate gradient solver in PyTorch using JVPs and VJPs, whereas in our system we explicitly materialize the Jacobian and use cuSPARSE to form $J^{\mathsf{T}}J$ and the right-hand side. We optimize the PyTorch implementation to the best of our ability; nevertheless, the JVP and VJP evaluations dominate the computation, accounting for approximately $99\%$ of the end-to-end time of each GN iteration.

\begin{wraptable}[9]{r}{0.45\linewidth}
    \centering
    \vspace{-2pt}
    \caption{Runtime breakdown for 60 GN iterations on 253k-face mesh.}
    \label{tab:polyvector_breakdown}
    \vspace{-6pt}
    \begin{tabular}{l l}
        \hline
        Stage                  & T (ms) \\
        \hline
        \textbf{Construct $J$} & 304.4  \\
        \hline
        \textbf{Assemble $J$}  & 98.7   \\
        Line Search            & 16.7   \\
        Linear Solver          & 8089.8 \\
        \hline
    \end{tabular}
    \vspace{-0.75em}
\end{wraptable}

In our system, we pay a modest upfront cost to construct the sparse Jacobian from the problem definition. Table~\ref{tab:polyvector_breakdown} reports a detailed runtime breakdown for 60 GN iterations on a mesh with 253k faces. The differentiation and assembly of $J$ together account for a small fraction of the total runtime, while the linear solve dominates the computation. End-to-end, our implementation completes the full optimization in \textbf{8.22 seconds}, compared to \textbf{377.49 seconds} for PyTorch (Figure~\ref{fig:polyvector}).

This application shows that explicitly materializing sparse Jacobians can be a practical and advantageous strategy for large-scale nonlinear geometry problems, provided the system can construct them efficiently. By exposing the full structure of $J$, our approach enables direct reuse of mature sparse linear algebra solvers rather than relying on repeated JVP/VJP evaluations. As a concrete example, replacing conjugate gradients with the cuDSS direct solver~\cite{NVIDIA:2025:cudss} reduces the runtime to 2.81 seconds, yielding an additional 2.9$\times$ speedup for the same problem shown in Figure~\ref{fig:polyvector}.

\subsection{Manifold Optimization}
\label{sec:manifold}
\begin{figure}
    \centering
    \includegraphics[width=0.48\textwidth]{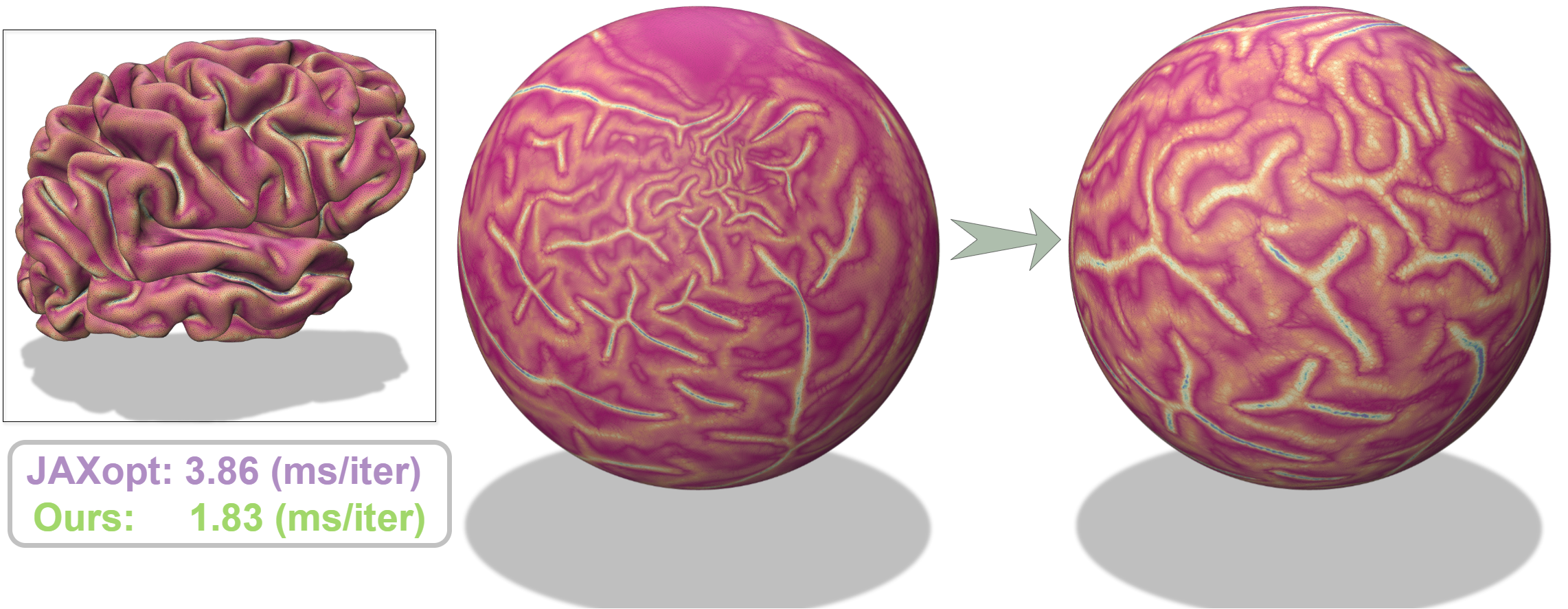}
    \caption{Starting from an initial embedding on the unit sphere (middle), we optimize vertex positions in tangent space to reduce distortion while preventing triangle flips, producing a smoother and more uniform parameterization (right). Using L-BFGS optimizer, and compared against JIT-compiled code from JAXopt~\cite{Blondel:2022:EAM}, our system achieves a \textbf2.1$\times$ speedup on a mesh with approximately 350k faces.}
    \label{fig:manifold_brain}
\end{figure}
We implement spherical parameterization of a genus-0 triangle mesh by embedding it onto the unit sphere $\mathbb{S}^2 \subset \mathbb{R}^3$ while minimizing distortion and avoiding triangle flips~\cite{Schmidt:2022:TAD}. Each vertex is represented as a point on $\mathbb{S}^2$ and optimization is performed in the 2D tangent space at each vertex. A retraction operator $R: \mathbb{S}^2 \times \mathbb{R}^2 \rightarrow \mathbb{S}^2 $ maps updates back to the manifold. At each iteration, we compute the derivatives in tangent space, run one step of the solver, and update the tangent basis.

The objective combines two terms: (1)~a barrier function that penalizes flipped triangles, and (2)~a stretch penalty encouraging uniform edge lengths. Formally, the objective is:
\begin{align}
    \min_{x \in \mathbb{R}^{2|V|}} \;
    \sum_{(i,j,k) \in F} \big[
     & -\log \det ( [p_i, p_j, p_k] ) \nonumber                    \\
     & + \| p_i - p_j \|^2 + \| p_j - p_k \|^2 + \| p_k - p_i \|^2
        \big]
        \label{eq:manifold}
\end{align}
where each $p_i = R(s_i, x_i) \in \mathbb{S}^2$ is obtained by retracting a 2D tangent vector $x_i = \begin{bmatrix} x_{i,1} & x_{i,2} \end{bmatrix}^\top \in \mathbb{R}^2$ at base point $s_i \in \mathbb{S}^2$ using a local orthonormal basis $ b_{1,i}, b_{2,i} \in \mathbb{R}^3$. The retraction is defined as:
$$
    R(s_i, x_i) =
    \frac{
        s_i + x_{i,1} \cdot b_{1,i} + x_{i,2} \cdot b_{2,i}
    }{
        \left\| s_i + x_{i,1} \cdot b_{1,i} + x_{i,2} \cdot b_{2,i} \right\|
    }
$$

We optimize the energy using L-BFGS with backtracking line search (Figure~\ref{fig:manifold_brain}). For comparison, we use JAXopt~\cite{Blondel:2022:EAM}, a JAX-based optimization library that also supports L-BFGS with backtracking line search. All JAXopt computations, including differentiation and linear algebra operations, are fully \texttt{jit}-compiled. Using the same configuration, on a mesh with 350k faces, our system achieves 1.83~ms/iteration, compared to JAXopt's 3.86~ms/iteration, yielding a \textbf{2.1$\mathbf{\times}$ speedup}. While JAX may better optimize the vector operations in L-BFGS, differentiation remains the dominant cost. This result highlights that our system maintains strong performance even in first-order methods, despite not explicitly targeting aggressive kernel fusion or full linear algebra acceleration. Incorporating such techniques is a promising direction for future work. In the next application, we isolate differentiation performance using gradient descent.

\subsection{Area-Based Smoothing}
\label{sec:area}
This application demonstrates our system's performance in first-order optimization through a simple geometric task, i.e., minimizing the total surface area of a triangle mesh. Given vertex positions $x \in \mathbb{R}^{3|V|}$ and a triangular mesh with face set $\mathbb{F}$, the energy is defined as the surface area functional $$E(x) = \sum_{(i,j,k)\in F} A_{ijk}$$
where $A_{ijk} = \tfrac12 \| (x_j - x_i) \times (x_k - x_i) \|$ is the area of triangle $(i,j,k)$. This energy encourages adjacent vertices to reduce surface area, effectively smoothing geometric noise, corresponding to mean curvature flow of smooth surfaces. The gradient contribution of a triangle $(i,j,k)$ to its vertices is given by
$
\frac{\partial A_{ijk}}{\partial x_i} = \tfrac12 (x_j - x_k) \times n,
$
where $n$ is the unit normal of the triangle. The expressions for $\partial A_{ijk} / \partial x_j$ and $\partial A_{ijk} / \partial x_k$ follow by cyclic permutation of the vertex indices. The full gradient for a vertex $x_i$ is obtained by summing contributions from all incident faces.

We use gradient descent for this optimization as 
$$x_i \gets x_i - \lambda \frac{\partial E}{\partial x_i},$$
with step size $\lambda > 0$. Full user code is shown in Appendix~\ref{appendix:smoothing}. While simplified, this optimization is a forward Euler implementation of the mean curvature flow and exemplifies a common pattern of mesh-based optimization via assembling local, face-wise gradients and updating vertex positions accordingly.

\begin{figure}
    \centering
    \includegraphics[width=0.48\textwidth]{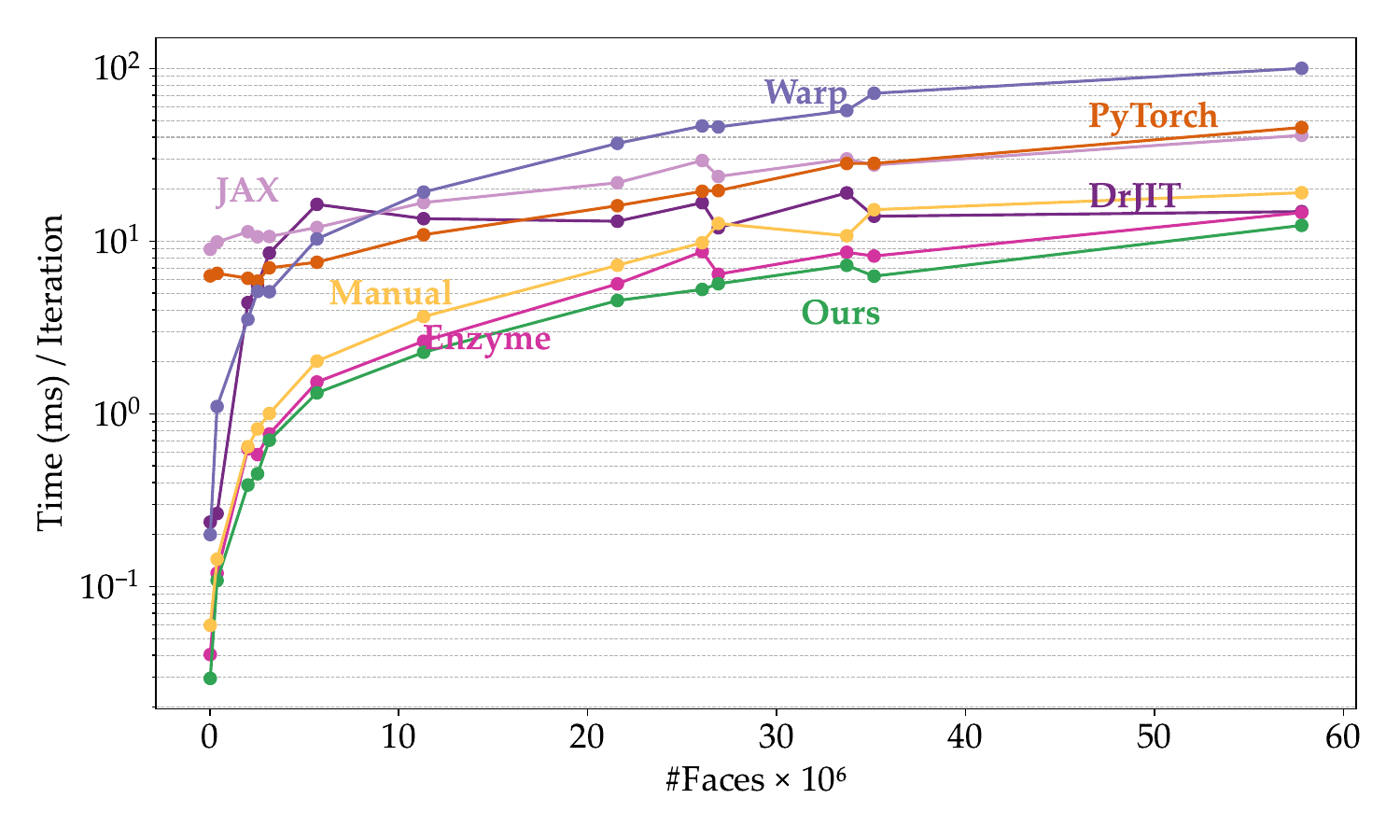}
    \caption{Scaling performance of our system compared to other GPU-based frameworks (PyTorch, Warp, Jax, EnzymeAD, and Dr.JIT), a manually implemented GPU baseline on area-based smoothing application which requires gradient computations only. Reported times are the average per gradient-descent step over 20 iterations.}
    \label{fig:smoothing_scale}
\end{figure}

We benchmark gradient computation against several AD frameworks, including: PyTorch and JAX (reverse-mode with JIT); Warp, EnzymeAD, and Dr.JIT (GPU-optimized reverse-mode); and a hand-coded baseline with closed-form gradients. Results in Figure~\ref{fig:smoothing_scale} show that our system delivers strong GPU performance with geometric mean speedup of \textbf{8.87$\mathbf{\times}$ over PyTorch}, \textbf{8.63$\mathbf{\times}$ over Warp}, \textbf{12.19$\mathbf{\times}$ over JAX}, \textbf{4.53$\mathbf{\times}$ over Dr.JIT}, \textbf{1.26$\mathbf{\times}$ over EnzymeAD}, and \textbf{1.71$\mathbf{\times}$} over handwritten gradient baseline. We implement the handwritten gradient baseline in RXMesh~\cite{Mahmoud:2021:RAG} so that this baseline can take advantage of the optimized memory layout in RXMesh. 

This result indicates that our system introduces negligible overhead and validates our use of atomic scatter for gradient accumulation for this application. Although atomics are often associated with contention-related slowdowns, mesh-based workloads typically exhibit low write conflicts. The manual implementation here uses a gather-based accumulation strategy (i.e., no atomics) and the fact that both approaches yield similar performance further supports the effectiveness of scatter in this context.

\subsection{Elastic Shell Simulation with Incremental Potential Contact}
\label{sec:elastic_sim}

We demonstrate a complex elastic simulation with several objects in motion and collision with each other (see Figure~\ref{fig:teaser}). The physical description of the scene consists of several energy terms that govern the simulation. Alongside the gravitational energy as in~\S\ref{sec:spring}, we use:
\begin{enumerate}
    \item \textit{Neo-Hookean Elasticity:} The material is modeled using Neo-Hookean strain energy density~\cite{Sifakis:2012:FSO} computed on the mesh elements as described in~\cite{Li:2024:PBS}.
    \item \textit{Discrete Bending Energy:} To maintain the global structural integrity of the mesh, we apply bending energy on the dihedral angles between adjacent triangles~\cite{Grinspun:2003:DS}. For a more robust implementation across different scales, we use the pre-computed area of the triangle as a factor instead of the height of the triangle~\cite{Tamstorf:2013:DBF}:
          $$ E_{\text{bend}}(\theta) = k_b \frac{A}{3} \left(\theta - \theta'\right)^2 h^2$$
          where $k_b$ is the bending stiffness, $A$ is the average area of the two triangles sharing the edge, $\theta$ is the current dihedral angle, $\theta'$ is the rest dihedral angle and $h$ is the time step.
    \item \textit{Inertial Energy:} To account for dynamics under implicit time integration, we include an inertial energy that penalizes deviations of vertex positions from their positions in the previous time step:
          $$E_{\text{inertia}}(x) = \frac{1}{2} m (x - x')^2$$
          where $m$ is the mass associated with each vertex, $x$ is the position of the vertex, and $x'$ is the position of the vertex in the previous time step.
    \item \textit{Box Barrier Energy:} To enforce spatial bounds, we apply a rescaled barrier energy~\cite{Li:2024:PBS} modeling the simulation box walls:
          $$E_{\text{box}}(d) = \frac{\kappa}{2} \hat{d} A_c (s - 1) \ln{\left(s\right)} h^2$$
          and $s = \frac{d}{\hat{d}}$, where $d$ is surface distance, $\hat{d}$ is activation distance, $\kappa$ is barrier stiffness, $A_c$ is average contact area, and $h$ is the time step. This energy comes into play when the distance $d$ is within the threshold $\hat{d}$.
    \item \textit{Contact Barrier Energy:} To prevent inter-object penetration, we add barrier energy terms for vertex-vertex (\texttt{Op:VV}) and vertex-face (\texttt{Op:FV}) contacts when elements approach within a threshold distance. The formulation $E_{\text{contact}(d)}$ matches the box barrier energy, with $A_c$ set to a small constant for vertex-vertex contacts and to face area for vertex-face contacts, and $d$ being the closest distance between the two participating mesh elements.

\end{enumerate}
The computation of the contact barrier energy requires finding the potential contact pairs of mesh elements, i.e., possible mesh elements undergoing collision. This search is accelerated using a BVH (adopted from \citet{Wald:2026:cuBQL}) and also depicts the ability to integrate acceleration data structures with our system.

We discretize the simulation in time and solve each time step using implicit Euler integration. Following~\cite{Li:2020:IPC}, we use a Newton solver with preconditioned conjugate gradient and a contact-aware line search to perform penetration-free descent steps.

\begin{figure}
    \centering
    \fbox{\includegraphics[width=0.28\linewidth]{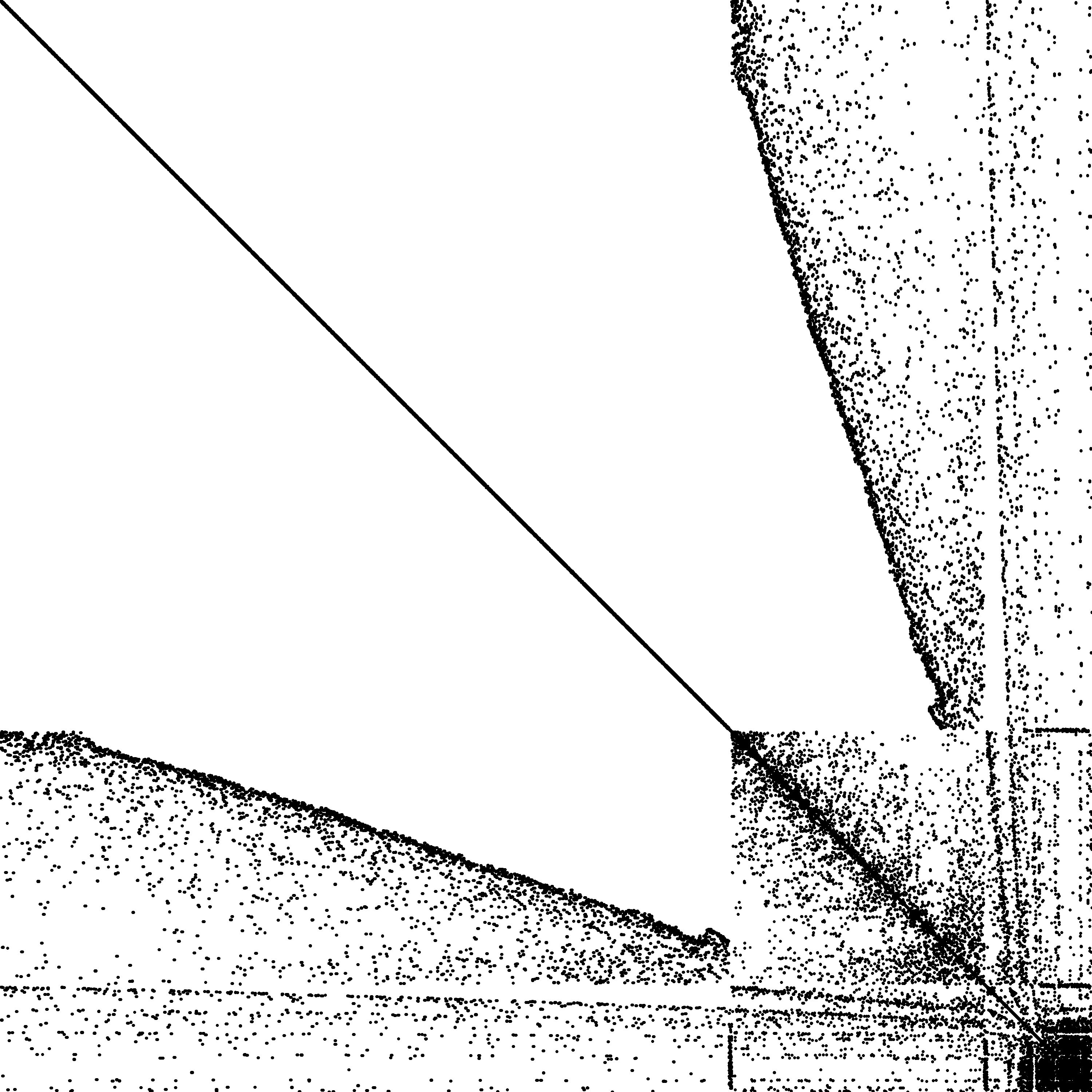}}
    \fbox{\includegraphics[width=0.28\linewidth]{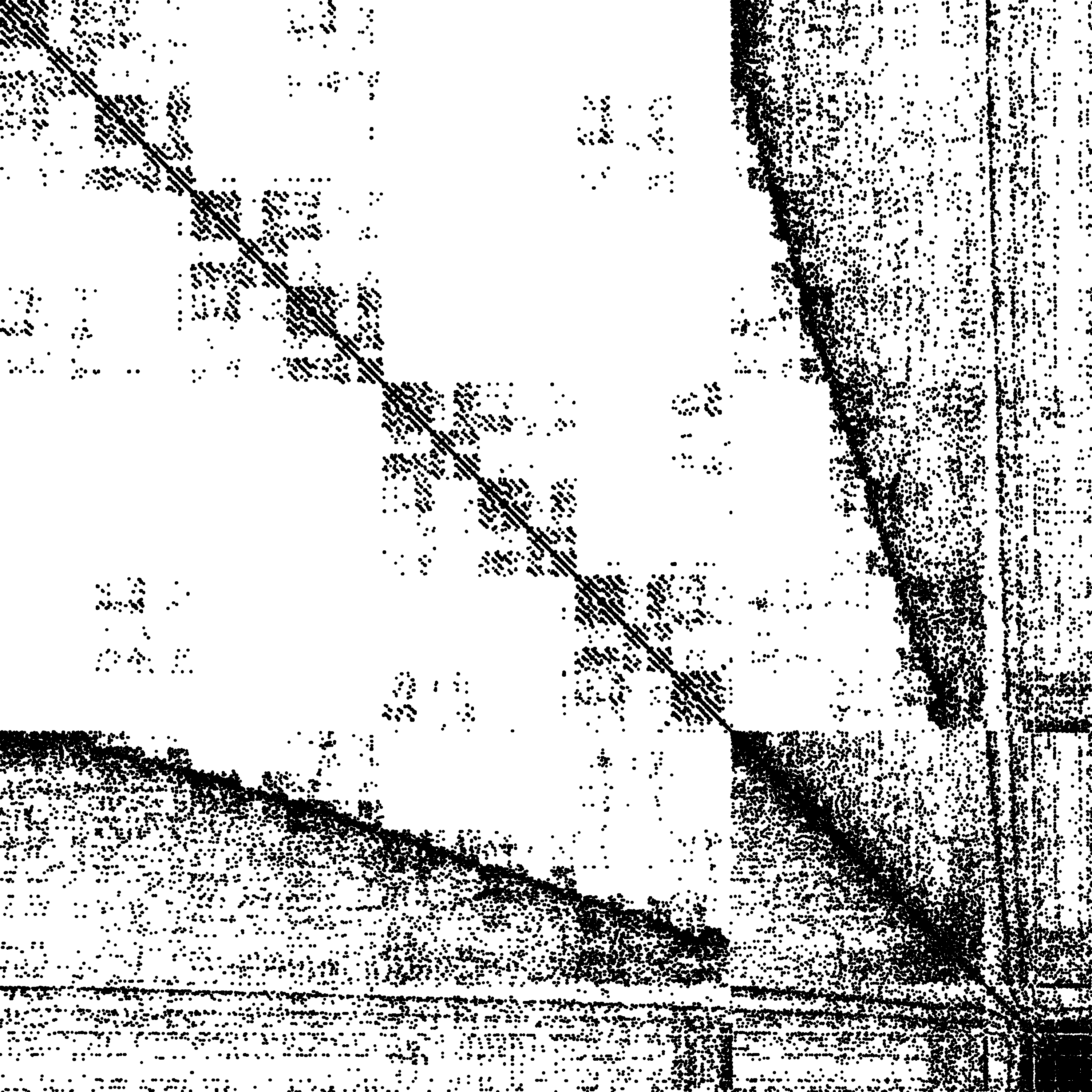}}
    \fbox{\includegraphics[width=0.28\linewidth]{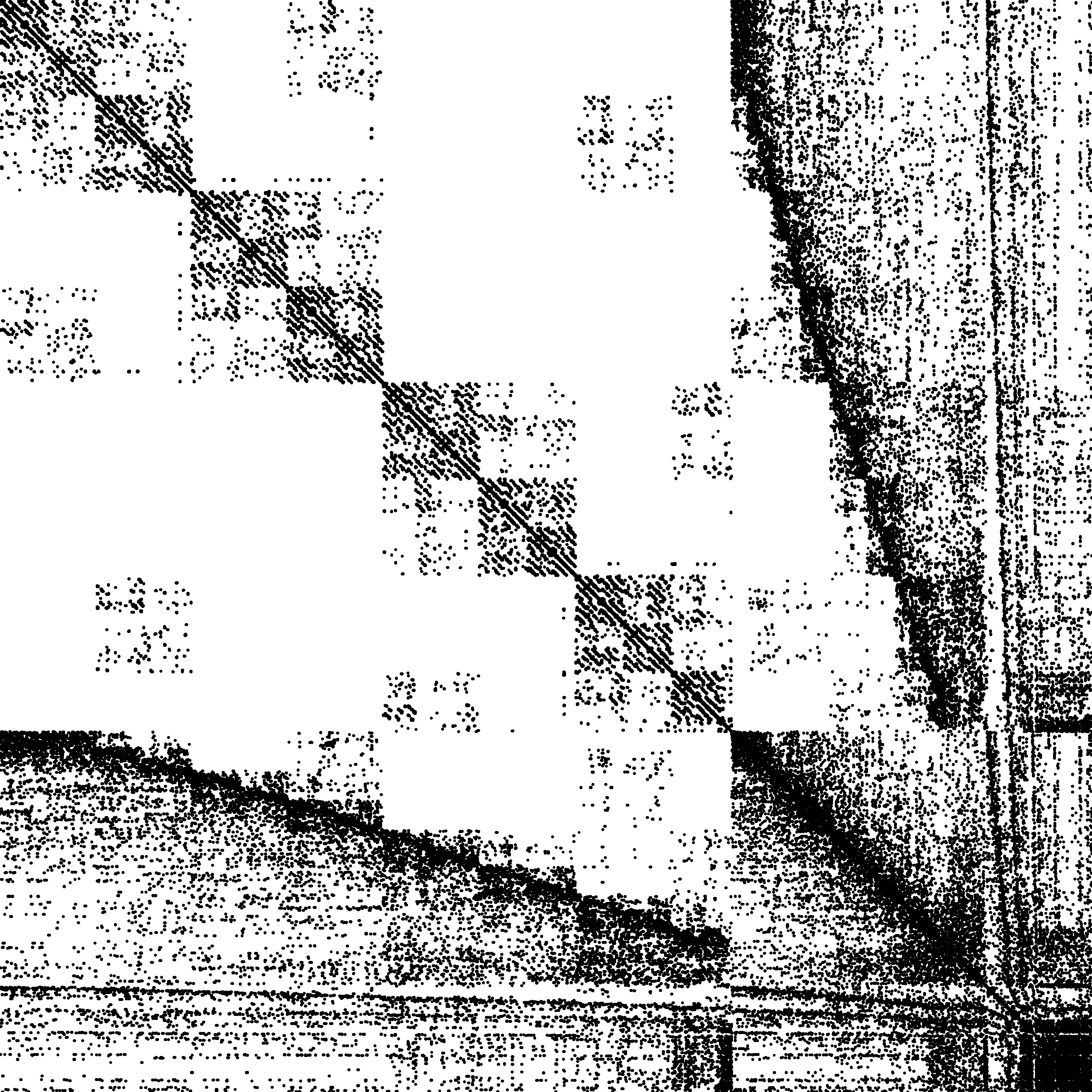}}
    \caption{Visualization of how the Sparse Hessian matrices evolve over time for three different time-steps of the large-scale elastic shell simulation depicted in Figure~\ref{fig:teaser}.}
    \label{fig:teaser_hessian}
\end{figure}

\begin{wraptable}[11]{r}{0.5\linewidth}
    \centering
    \vspace{-10pt}
    \caption{Per-frame runtime breakdown of the scene in Figure~\ref{fig:teaser} }
    \label{tab:elastic_timing}
    \vspace{-6pt}
    \begin{tabular}{l l}
        \hline
        Stage                   & Time (s)               \\
        \hline
        Contact Detect          & 97.9 (49.1\%)          \\
        Linear Solver           & 76.2 (38.2\%)          \\
        \textbf{Energy Eval}    & \textbf{21.1 (10.5\%)} \\
        \textbf{Hessian Update} & \textbf{3.3  (1.7\%)}  \\
        Line Search             & 0.6   (0.27\%)         \\
        Misc                    & 0.06  (0.03\%)         \\
        \hline
    \end{tabular}
    \vspace{-0.75em}
\end{wraptable}
We evaluate performance on the large-scale contact scenario shown in Figure~\ref{fig:teaser}. The scene contains 2.1M vertices and 4.12M faces across all objects with the number of active contact interactions dynamically changing over time and peaking at $0.34$M vertex--vertex and $1.85$M vertex--face pairs which impacts the sparsity of the Hessian over time (see Figure~\ref{fig:teaser_hessian}). Table~\ref{tab:elastic_timing} reports a detailed per-time step breakdown---separating contact pair generation (BVH construction and traversal), linear solves, derivative evaluation, Hessian updates, and line search.

Here, the computational cost is dominated by contact handling and the linear solver which together account for approximately 87.3\% of total runtime. Despite the highly dynamic and irregular sparsity induced by contact, automatic differentiation and sparse assembly remain negligible, contributing only 12.2\% of runtime. This confirms that our locality-aware design effectively removes differentiation from the critical path even in regimes where sparsity patterns change. 

\section{Conclusion}
\label{sec:conclusion}
Derivatives, in particular second-order derivatives, are often treated as prohibitively expensive in large-scale simulation and geometry-processing pipelines. As a result, many practical systems either avoid Hessians altogether or rely on first-order methods even when faster and more robust second-order solvers would be preferable. This work shows that, for mesh-based problems, this trade-off is not fundamental. By exploiting the locality and sparsity inherent in mesh energies, we demonstrate that sparse gradients, Jacobians, and Hessians can be computed efficiently and at scale.

An important implication of this result is that second-order optimization need no longer be confined to small or moderately sized meshes. In our experiments, derivative evaluation becomes a minor fraction of total runtime even for large problems, shifting the dominant cost to linear solvers and collision detection. This opens the door to routinely using Newton in interactive deformation, physical simulation, and inverse problems at resolutions that were previously impractical. More broadly, it suggests that locality-aware differentiation can serve as a viable foundation for a new generation of GPU numerical solvers for unstructured problems.

Looking forward, several directions are particularly promising. First, extending the system to volumetric discretizations, such as tetrahedral meshes, would enable efficient differentiation for a broader class of solid mechanics and multiphysics simulations. Second, while our current design launches each energy term as a separate kernel, a compiler that can fuse multiple local terms into a single kernel would further reduce memory traffic and could unlock additional performance gains. Finally, integrating dynamic mesh connectivity through patch-level updates, as well as supporting wider $k$-ring stencils and composed operators (e.g., bi-Laplacians), would broaden the expressiveness of the programming model.

\begin{acks}
\label{sec:ack}
The MIT Geometric Data Processing Group acknowledges the generous support of National Science Foundation grants IIS2335492 and OAC2403239, from the CSAIL Future of Data and FinTechAI programs, from the MIT--IBM Watson AI Laboratory, from the Wistron Corporation, from the MIT Generative AI Impact Consortium, from the Toyota--CSAIL Joint Research Center, and from Schmidt Sciences.
\end{acks}

\bibliographystyle{ACM-Reference-Format}
\bibliography{ahmed_ref}


\begin{thebibliography}{42}


\ifx \showCODEN    \undefined \def \showCODEN     #1{\unskip}     \fi
\ifx \showISBNx    \undefined \def \showISBNx     #1{\unskip}     \fi
\ifx \showISBNxiii \undefined \def \showISBNxiii  #1{\unskip}     \fi
\ifx \showISSN     \undefined \def \showISSN      #1{\unskip}     \fi
\ifx \showLCCN     \undefined \def \showLCCN      #1{\unskip}     \fi
\ifx \shownote     \undefined \def \shownote      #1{#1}          \fi
\ifx \showarticletitle \undefined \def \showarticletitle #1{#1}   \fi
\ifx \showURL      \undefined \def \showURL       {\relax}        \fi
\providecommand\bibfield[2]{#2}
\providecommand\bibinfo[2]{#2}
\providecommand\natexlab[1]{#1}
\providecommand\showeprint[2][]{arXiv:#2}

\bibitem[Ansel et~al\mbox{.}(2024)]%
        {Ansel:2024:P2F}
\bibfield{author}{\bibinfo{person}{Jason Ansel}, \bibinfo{person}{Edward Yang},
  \bibinfo{person}{Horace He}, \bibinfo{person}{Natalia Gimelshein},
  \bibinfo{person}{Animesh Jain}, \bibinfo{person}{Michael Voznesensky},
  \bibinfo{person}{Bin Bao}, \bibinfo{person}{Peter Bell},
  \bibinfo{person}{David Berard}, \bibinfo{person}{Evgeni Burovski},
  \bibinfo{person}{Geeta Chauhan}, \bibinfo{person}{Anjali Chourdia},
  \bibinfo{person}{Will Constable}, \bibinfo{person}{Alban Desmaison},
  \bibinfo{person}{Zachary DeVito}, \bibinfo{person}{Elias Ellison},
  \bibinfo{person}{Will Feng}, \bibinfo{person}{Jiong Gong},
  \bibinfo{person}{Michael Gschwind}, \bibinfo{person}{Brian Hirsh},
  \bibinfo{person}{Sherlock Huang}, \bibinfo{person}{Kshiteej Kalambarkar},
  \bibinfo{person}{Laurent Kirsch}, \bibinfo{person}{Michael Lazos},
  \bibinfo{person}{Mario Lezcano}, \bibinfo{person}{Yanbo Liang},
  \bibinfo{person}{Jason Liang}, \bibinfo{person}{Yinghai Lu},
  \bibinfo{person}{CK Luk}, \bibinfo{person}{Bert Maher},
  \bibinfo{person}{Yunjie Pan}, \bibinfo{person}{Christian Puhrsch},
  \bibinfo{person}{Matthias Reso}, \bibinfo{person}{Mark Saroufim},
  \bibinfo{person}{Marcos~Yukio Siraichi}, \bibinfo{person}{Helen Suk},
  \bibinfo{person}{Michael Suo}, \bibinfo{person}{Phil Tillet},
  \bibinfo{person}{Eikan Wang}, \bibinfo{person}{Xiaodong Wang},
  \bibinfo{person}{William Wen}, \bibinfo{person}{Shunting Zhang},
  \bibinfo{person}{Xu Zhao}, \bibinfo{person}{Keren Zhou},
  \bibinfo{person}{Richard Zou}, \bibinfo{person}{Ajit Mathews},
  \bibinfo{person}{Gregory Chanan}, \bibinfo{person}{Peng Wu}, {and}
  \bibinfo{person}{Soumith Chintala}.} \bibinfo{year}{2024}\natexlab{}.
\newblock \showarticletitle{{P}y{T}orch 2: Faster Machine Learning Through
  Dynamic Python Bytecode Transformation and Graph Compilation}. In
  \bibinfo{booktitle}{\emph{29th ACM International Conference on Architectural
  Support for Programming Languages and Operating Systems, Volume 2 (ASPLOS
  '24)}}. \bibinfo{publisher}{ACM}.
\newblock
\href{https://doi.org/10.1145/3620665.3640366}{doi:\nolinkurl{10.1145/3620665.3640366}}


\bibitem[Blondel et~al\mbox{.}(2022)]%
        {Blondel:2022:EAM}
\bibfield{author}{\bibinfo{person}{Mathieu Blondel}, \bibinfo{person}{Quentin
  Berthet}, \bibinfo{person}{Marco Cuturi}, \bibinfo{person}{Roy Frostig},
  \bibinfo{person}{Stephan Hoyer}, \bibinfo{person}{Felipe Llinares-L{\'o}pez},
  \bibinfo{person}{Fabian Pedregosa}, {and} \bibinfo{person}{Jean-Philippe
  Vert}.} \bibinfo{year}{2022}\natexlab{}.
\newblock \showarticletitle{Efficient and Modular Implicit Differentiation}. In
  \bibinfo{booktitle}{\emph{Proceedings of the 36th International Conference on
  Neural Information Processing Systems}} \emph{(\bibinfo{series}{NIPS '22},
  Vol.~\bibinfo{volume}{35})}. \bibinfo{publisher}{Curran Associates, Inc.},
  Article \bibinfo{articleno}{378}, \bibinfo{numpages}{13}~pages.
\newblock
\showISBNx{9781713871088}
\href{https://doi.org/10.5555/3600270.3600648}{doi:\nolinkurl{10.5555/3600270.3600648}}


\bibitem[Bradbury et~al\mbox{.}(2018)]%
        {Bradbury:2018:JCT}
\bibfield{author}{\bibinfo{person}{James Bradbury}, \bibinfo{person}{Roy
  Frostig}, \bibinfo{person}{Peter Hawkins}, \bibinfo{person}{Matthew~James
  Johnson}, \bibinfo{person}{Chris Leary}, \bibinfo{person}{Dougal Maclaurin},
  \bibinfo{person}{George Necula}, \bibinfo{person}{Adam Paszke},
  \bibinfo{person}{Jake Vander{P}las}, \bibinfo{person}{Skye
  Wanderman-{M}ilne}, {and} \bibinfo{person}{Qiao Zhang}.}
  \bibinfo{year}{2018}\natexlab{}.
\newblock \bibinfo{booktitle}{\emph{{JAX}: composable transformations of
  {P}ython+{N}um{P}y programs}}.
\newblock
\urldef\tempurl%
\url{http://github.com/jax-ml/jax}
\showURL{%
\tempurl}


\bibitem[Brisson(1989)]%
        {Brisson:1989:RGS}
\bibfield{author}{\bibinfo{person}{Erik Brisson}.}
  \bibinfo{year}{1989}\natexlab{}.
\newblock \showarticletitle{Representing Geometric Structures in $d$
  Dimensions: Topology and Order}. In \bibinfo{booktitle}{\emph{Proceedings of
  the Fifth Annual Symposium on Computational Geometry}} (Saarbruchen, West
  Germany) \emph{(\bibinfo{series}{SCG '89})}. \bibinfo{publisher}{Association
  for Computing Machinery}, \bibinfo{address}{New York, NY, USA},
  \bibinfo{pages}{218--227}.
\newblock
\showISBNx{0897913183}
\href{https://doi.org/10.1145/73833.73858}{doi:\nolinkurl{10.1145/73833.73858}}


\bibitem[Devito et~al\mbox{.}(2017)]%
        {Devito:2017:OAD}
\bibfield{author}{\bibinfo{person}{Zachary Devito}, \bibinfo{person}{Michael
  Mara}, \bibinfo{person}{Michael Zollh\"{o}fer}, \bibinfo{person}{Gilbert
  Bernstein}, \bibinfo{person}{Jonathan Ragan-Kelley},
  \bibinfo{person}{Christian Theobalt}, \bibinfo{person}{Pat Hanrahan},
  \bibinfo{person}{Matthew Fisher}, {and} \bibinfo{person}{Matthias Niessner}.}
  \bibinfo{year}{2017}\natexlab{}.
\newblock \showarticletitle{Opt: A Domain Specific Language for Non-Linear
  Least Squares Optimization in Graphics and Imaging}.
\newblock \bibinfo{journal}{\emph{ACM Trans. Graph.}} \bibinfo{volume}{36},
  \bibinfo{number}{5}, Article \bibinfo{articleno}{171} (\bibinfo{date}{Oct.}
  \bibinfo{year}{2017}), \bibinfo{numpages}{27}~pages.
\newblock
\showISSN{0730-0301}
\href{https://doi.org/10.1145/3132188}{doi:\nolinkurl{10.1145/3132188}}


\bibitem[Diamanti et~al\mbox{.}(2015)]%
        {Diamanti:2015:IPF}
\bibfield{author}{\bibinfo{person}{Olga Diamanti}, \bibinfo{person}{Amir
  Vaxman}, \bibinfo{person}{Daniele Panozzo}, {and} \bibinfo{person}{Olga
  Sorkine-Hornung}.} \bibinfo{year}{2015}\natexlab{}.
\newblock \showarticletitle{Integrable PolyVector Fields}.
\newblock \bibinfo{journal}{\emph{ACM Trans. Graph.}} \bibinfo{volume}{34},
  \bibinfo{number}{4}, Article \bibinfo{articleno}{38} (\bibinfo{date}{July}
  \bibinfo{year}{2015}), \bibinfo{numpages}{12}~pages.
\newblock
\showISSN{0730-0301}
\href{https://doi.org/10.1145/2766906}{doi:\nolinkurl{10.1145/2766906}}


\bibitem[Fern\'{a}ndez-Fern\'{a}ndez et~al\mbox{.}(2025)]%
        {Fernandez:2025:SEB}
\bibfield{author}{\bibinfo{person}{Jos\'{e}~Antonio
  Fern\'{a}ndez-Fern\'{a}ndez}, \bibinfo{person}{Fabian L\"{o}schner},
  \bibinfo{person}{Lukas Westhofen}, \bibinfo{person}{Andreas Longva}, {and}
  \bibinfo{person}{Jan Bender}.} \bibinfo{year}{2025}\natexlab{}.
\newblock \showarticletitle{{SymX}: Energy-based Simulation from Symbolic
  Expressions}.
\newblock \bibinfo{journal}{\emph{ACM Trans. Graph.}} \bibinfo{volume}{45},
  \bibinfo{number}{1}, Article \bibinfo{articleno}{5} (\bibinfo{date}{Oct.}
  \bibinfo{year}{2025}), \bibinfo{numpages}{19}~pages.
\newblock
\showISSN{0730-0301}
\href{https://doi.org/10.1145/3764928}{doi:\nolinkurl{10.1145/3764928}}


\bibitem[Griewank et~al\mbox{.}(1996)]%
        {Griewank:1996:A7A}
\bibfield{author}{\bibinfo{person}{Andreas Griewank}, \bibinfo{person}{David
  Juedes}, {and} \bibinfo{person}{Jean Utke}.} \bibinfo{year}{1996}\natexlab{}.
\newblock \showarticletitle{Algorithm 755: ADOL-C: a package for the automatic
  differentiation of algorithms written in C/C++}.
\newblock  \bibinfo{volume}{22}, \bibinfo{number}{2} (\bibinfo{date}{June}
  \bibinfo{year}{1996}), \bibinfo{pages}{131--167}.
\newblock
\showISSN{0098-3500}
\href{https://doi.org/10.1145/229473.229474}{doi:\nolinkurl{10.1145/229473.229474}}


\bibitem[Griewank and Walther(2008)]%
        {Griewank:2008:EDP}
\bibfield{author}{\bibinfo{person}{Andreas Griewank} {and}
  \bibinfo{person}{Andrea Walther}.} \bibinfo{year}{2008}\natexlab{}.
\newblock \bibinfo{booktitle}{\emph{Evaluating Derivatives: Principles and
  Techniques of Algorithmic Differentiation} (\bibinfo{edition}{second} ed.)}.
\newblock \bibinfo{publisher}{Society for Industrial and Applied Mathematics},
  \bibinfo{address}{USA}.
\newblock
\showISBNx{0898716594}
\href{https://doi.org/10.5555/1455489}{doi:\nolinkurl{10.5555/1455489}}


\bibitem[Grinspun et~al\mbox{.}(2003)]%
        {Grinspun:2003:DS}
\bibfield{author}{\bibinfo{person}{Eitan Grinspun}, \bibinfo{person}{Anil~N.
  Hirani}, \bibinfo{person}{Mathieu Desbrun}, {and} \bibinfo{person}{Peter
  Schr\"{o}der}.} \bibinfo{year}{2003}\natexlab{}.
\newblock \showarticletitle{Discrete Shells}. In
  \bibinfo{booktitle}{\emph{Proceedings of the 2003 ACM SIGGRAPH/Eurographics
  Symposium on Computer Animation}} (San Diego, California)
  \emph{(\bibinfo{series}{SCA '03})}. \bibinfo{publisher}{Eurographics
  Association}, \bibinfo{address}{Goslar, DEU}, \bibinfo{pages}{62--67}.
\newblock
\showISBNx{1581136595}


\bibitem[Herholz et~al\mbox{.}(2024)]%
        {Herholz:2024:AMB}
\bibfield{author}{\bibinfo{person}{Philipp Herholz}, \bibinfo{person}{Tuur
  Stuyck}, {and} \bibinfo{person}{Ladislav Kavan}.}
  \bibinfo{year}{2024}\natexlab{}.
\newblock \showarticletitle{A Mesh-based Simulation Framework using Automatic
  Code Generation}.
\newblock \bibinfo{journal}{\emph{ACM Trans. Graph.}} \bibinfo{volume}{43},
  \bibinfo{number}{6}, Article \bibinfo{articleno}{215} (\bibinfo{date}{Nov.}
  \bibinfo{year}{2024}), \bibinfo{numpages}{17}~pages.
\newblock
\showISSN{0730-0301}
\href{https://doi.org/10.1145/3687986}{doi:\nolinkurl{10.1145/3687986}}


\bibitem[Herholz et~al\mbox{.}(2022)]%
        {Herholz:2022:SSC}
\bibfield{author}{\bibinfo{person}{Philipp Herholz}, \bibinfo{person}{Xuan
  Tang}, \bibinfo{person}{Teseo Schneider}, \bibinfo{person}{Shoaib Kamil},
  \bibinfo{person}{Daniele Panozzo}, {and} \bibinfo{person}{Olga
  Sorkine-Hornung}.} \bibinfo{year}{2022}\natexlab{}.
\newblock \showarticletitle{Sparsity-Specific Code Optimization using
  Expression Trees}.
\newblock \bibinfo{journal}{\emph{ACM Trans. Graph.}} \bibinfo{volume}{41},
  \bibinfo{number}{5}, Article \bibinfo{articleno}{175} (\bibinfo{date}{May}
  \bibinfo{year}{2022}), \bibinfo{numpages}{19}~pages.
\newblock
\showISSN{0730-0301}
\href{https://doi.org/10.1145/3520484}{doi:\nolinkurl{10.1145/3520484}}


\bibitem[Huang et~al\mbox{.}(2024)]%
        {Huang:2024:DSF}
\bibfield{author}{\bibinfo{person}{Zizhou Huang}, \bibinfo{person}{Davi~Colli
  Tozoni}, \bibinfo{person}{Arvi Gjoka}, \bibinfo{person}{Zachary Ferguson},
  \bibinfo{person}{Teseo Schneider}, \bibinfo{person}{Daniele Panozzo}, {and}
  \bibinfo{person}{Denis Zorin}.} \bibinfo{year}{2024}\natexlab{}.
\newblock \showarticletitle{Differentiable solver for time-dependent
  deformation problems with contact}.
\newblock \bibinfo{journal}{\emph{ACM Trans. Graph.}} \bibinfo{volume}{43},
  \bibinfo{number}{3}, Article \bibinfo{articleno}{31} (\bibinfo{date}{May}
  \bibinfo{year}{2024}), \bibinfo{numpages}{30}~pages.
\newblock
\showISSN{0730-0301}
\href{https://doi.org/10.1145/3657648}{doi:\nolinkurl{10.1145/3657648}}


\bibitem[Jakob(2019)]%
        {Jakob:2019:ESV}
\bibfield{author}{\bibinfo{person}{Wenzel Jakob}.}
  \bibinfo{year}{2019}\natexlab{}.
\newblock \bibinfo{title}{Enoki: structured vectorization and differentiation
  on modern processor architectures}.
\newblock
\newblock
\shownote{https://github.com/mitsuba-renderer/enoki}.


\bibitem[Jakob et~al\mbox{.}(2022)]%
        {Jakob:2022:DAJ}
\bibfield{author}{\bibinfo{person}{Wenzel Jakob},
  \bibinfo{person}{S\'{e}bastien Speierer}, \bibinfo{person}{Nicolas Roussel},
  {and} \bibinfo{person}{Delio Vicini}.} \bibinfo{year}{2022}\natexlab{}.
\newblock \showarticletitle{{DR.JIT}: a just-in-time compiler for
  differentiable rendering}.
\newblock \bibinfo{journal}{\emph{ACM Trans. Graph.}} \bibinfo{volume}{41},
  \bibinfo{number}{4}, Article \bibinfo{articleno}{124} (\bibinfo{date}{July}
  \bibinfo{year}{2022}), \bibinfo{numpages}{19}~pages.
\newblock
\showISSN{0730-0301}
\href{https://doi.org/10.1145/3528223.3530099}{doi:\nolinkurl{10.1145/3528223.3530099}}


\bibitem[Kim and Eberle(2020)]%
        {Kim:2020:DDI}
\bibfield{author}{\bibinfo{person}{Theodore Kim} {and} \bibinfo{person}{David
  Eberle}.} \bibinfo{year}{2020}\natexlab{}.
\newblock \showarticletitle{Dynamic deformables: implementation and production
  practicalities}. In \bibinfo{booktitle}{\emph{ACM SIGGRAPH 2020 Courses}}
  (Virtual Event, USA) \emph{(\bibinfo{series}{SIGGRAPH '20})}.
  \bibinfo{publisher}{Association for Computing Machinery},
  \bibinfo{address}{New York, NY, USA}, Article \bibinfo{articleno}{23},
  \bibinfo{numpages}{182}~pages.
\newblock
\showISBNx{9781450379724}
\href{https://doi.org/10.1145/3388769.3407490}{doi:\nolinkurl{10.1145/3388769.3407490}}


\bibitem[Kim and Eberle(2022)]%
        {Kim:2022:DDI}
\bibfield{author}{\bibinfo{person}{Theodore Kim} {and} \bibinfo{person}{David
  Eberle}.} \bibinfo{year}{2022}\natexlab{}.
\newblock \showarticletitle{Dynamic deformables: implementation and production
  practicalities (now with code!)}. In \bibinfo{booktitle}{\emph{ACM SIGGRAPH
  2022 Courses}} (Vancouver, British Columbia, Canada)
  \emph{(\bibinfo{series}{SIGGRAPH '22})}. \bibinfo{publisher}{Association for
  Computing Machinery}, \bibinfo{address}{New York, NY, USA}, Article
  \bibinfo{articleno}{7}, \bibinfo{numpages}{259}~pages.
\newblock
\showISBNx{9781450393621}
\href{https://doi.org/10.1145/3532720.3535628}{doi:\nolinkurl{10.1145/3532720.3535628}}


\bibitem[Li et~al\mbox{.}(2020)]%
        {Li:2020:IPC}
\bibfield{author}{\bibinfo{person}{Minchen Li}, \bibinfo{person}{Zachary
  Ferguson}, \bibinfo{person}{Teseo Schneider}, \bibinfo{person}{Timothy
  Langlois}, \bibinfo{person}{Denis Zorin}, \bibinfo{person}{Daniele Panozzo},
  \bibinfo{person}{Chenfanfu Jiang}, {and} \bibinfo{person}{Danny~M. Kaufman}.}
  \bibinfo{year}{2020}\natexlab{}.
\newblock \showarticletitle{Incremental Potential Contact: intersection-and
  inversion-free, large-deformation dynamics}.
\newblock \bibinfo{journal}{\emph{ACM Trans. Graph.}} \bibinfo{volume}{39},
  \bibinfo{number}{4}, Article \bibinfo{articleno}{49} (\bibinfo{date}{Aug.}
  \bibinfo{year}{2020}), \bibinfo{numpages}{20}~pages.
\newblock
\showISSN{0730-0301}
\href{https://doi.org/10.1145/3386569.3392425}{doi:\nolinkurl{10.1145/3386569.3392425}}


\bibitem[Li et~al\mbox{.}(2024)]%
        {Li:2024:PBS}
\bibfield{author}{\bibinfo{person}{Minchen Li}, \bibinfo{person}{Chenfanfu
  Jiang}, {and} \bibinfo{person}{Zhaofeng Luo}.}
  \bibinfo{year}{2024}\natexlab{}.
\newblock \bibinfo{booktitle}{\emph{Physics-Based Simulation}}.
\newblock
\urldef\tempurl%
\url{https://phys-sim-book.github.io/}
\showURL{%
\tempurl}


\bibitem[Macklin(2022)]%
        {Macklin:2022:WAH}
\bibfield{author}{\bibinfo{person}{Miles Macklin}.}
  \bibinfo{year}{2022}\natexlab{}.
\newblock \bibinfo{title}{Warp: A High-performance Python Framework for GPU
  Simulation and Graphics}.
\newblock \bibinfo{howpublished}{\url{https://github.com/nvidia/warp}}.
\newblock
\newblock
\shownote{NVIDIA GPU Technology Conference (GTC)}.


\bibitem[Mahmoud et~al\mbox{.}(2021)]%
        {Mahmoud:2021:RAG}
\bibfield{author}{\bibinfo{person}{Ahmed~H. Mahmoud},
  \bibinfo{person}{Serban~D. Porumbescu}, {and} \bibinfo{person}{John~D.
  Owens}.} \bibinfo{year}{2021}\natexlab{}.
\newblock \showarticletitle{{RXM}esh: A {GPU} Mesh Data Structure}.
\newblock \bibinfo{journal}{\emph{ACM Transactions on Graphics}}
  \bibinfo{volume}{40}, \bibinfo{number}{4}, Article \bibinfo{articleno}{104}
  (\bibinfo{date}{Aug.} \bibinfo{year}{2021}), \bibinfo{numpages}{16}~pages.
\newblock
\href{https://doi.org/10.1145/3450626.3459748}{doi:\nolinkurl{10.1145/3450626.3459748}}


\bibitem[Mahmoud et~al\mbox{.}(2025)]%
        {Mahmoud:2025:DMP}
\bibfield{author}{\bibinfo{person}{Ahmed~H. Mahmoud},
  \bibinfo{person}{Serban~D. Porumbescu}, {and} \bibinfo{person}{John~D.
  Owens}.} \bibinfo{year}{2025}\natexlab{}.
\newblock \showarticletitle{Dynamic Mesh Processing on the {GPU}}.
\newblock \bibinfo{journal}{\emph{ACM Transactions on Graphics}}
  \bibinfo{volume}{44}, \bibinfo{number}{4}, Article \bibinfo{articleno}{136}
  (\bibinfo{date}{July} \bibinfo{year}{2025}), \bibinfo{numpages}{19}~pages.
\newblock
\showISSN{0730-0301}
\href{https://doi.org/10.1145/3731162}{doi:\nolinkurl{10.1145/3731162}}


\bibitem[M\"{a}ntyl\"{a}(1988)]%
        {Mantyla:1988:ITS}
\bibfield{author}{\bibinfo{person}{M. M\"{a}ntyl\"{a}}.}
  \bibinfo{year}{1988}\natexlab{}.
\newblock \bibinfo{booktitle}{\emph{Introduction to Solid Modeling}}.
\newblock \bibinfo{publisher}{W. H. Freeman \& Co.}, \bibinfo{address}{New
  York, NY, USA}.
\newblock
\showISBNx{0-88175-108-1}


\bibitem[Mara et~al\mbox{.}(2021)]%
        {Mara:2021:TSF}
\bibfield{author}{\bibinfo{person}{Michael Mara}, \bibinfo{person}{Felix
  Heide}, \bibinfo{person}{Michael Zollh\"{o}fer}, \bibinfo{person}{Matthias
  Nie\ss{}ner}, {and} \bibinfo{person}{Pat Hanrahan}.}
  \bibinfo{year}{2021}\natexlab{}.
\newblock \showarticletitle{Thallo -- Scheduling for High-Performance
  Large-Scale Non-Linear Least-Squares Solvers}.
\newblock \bibinfo{journal}{\emph{ACM Trans. Graph.}} \bibinfo{volume}{40},
  \bibinfo{number}{5}, Article \bibinfo{articleno}{184} (\bibinfo{date}{Sept.}
  \bibinfo{year}{2021}), \bibinfo{numpages}{14}~pages.
\newblock
\showISSN{0730-0301}
\href{https://doi.org/10.1145/3453986}{doi:\nolinkurl{10.1145/3453986}}


\bibitem[Martins and Ning(2021)]%
        {Martins:2021:EDO}
\bibfield{author}{\bibinfo{person}{Joaquim R. R.~A. Martins} {and}
  \bibinfo{person}{Andrew Ning}.} \bibinfo{year}{2021}\natexlab{}.
\newblock \bibinfo{booktitle}{\emph{Engineering Design Optimization}}.
\newblock \bibinfo{publisher}{Cambridge University Press}.
\newblock
\showISBNx{9781108833417}


\bibitem[Moses et~al\mbox{.}(2021)]%
        {Moses:2021:RMA}
\bibfield{author}{\bibinfo{person}{William~S. Moses}, \bibinfo{person}{Valentin
  Churavy}, \bibinfo{person}{Ludger Paehler}, \bibinfo{person}{Jan
  H\"{u}ckelheim}, \bibinfo{person}{Sri Hari~Krishna Narayanan},
  \bibinfo{person}{Michel Schanen}, {and} \bibinfo{person}{Johannes Doerfert}.}
  \bibinfo{year}{2021}\natexlab{}.
\newblock \showarticletitle{Reverse-Mode Automatic Differentiation and
  Optimization of GPU Kernels via Enzyme}. In
  \bibinfo{booktitle}{\emph{Proceedings of the International Conference for
  High Performance Computing, Networking, Storage and Analysis}} (St. Louis,
  Missouri) \emph{(\bibinfo{series}{SC '21})}. \bibinfo{publisher}{Association
  for Computing Machinery}, \bibinfo{address}{New York, NY, USA}, Article
  \bibinfo{articleno}{61}, \bibinfo{numpages}{16}~pages.
\newblock
\showISBNx{9781450384421}
\href{https://doi.org/10.1145/3458817.3476165}{doi:\nolinkurl{10.1145/3458817.3476165}}


\bibitem[Moses et~al\mbox{.}(2022)]%
        {Moses:2022:SAD}
\bibfield{author}{\bibinfo{person}{William~S. Moses}, \bibinfo{person}{Sri
  Hari~Krishna Narayanan}, \bibinfo{person}{Ludger Paehler},
  \bibinfo{person}{Valentin Churavy}, \bibinfo{person}{Michel Schanen},
  \bibinfo{person}{Jan H\"{u}ckelheim}, \bibinfo{person}{Johannes Doerfert},
  {and} \bibinfo{person}{Paul Hovland}.} \bibinfo{year}{2022}\natexlab{}.
\newblock \showarticletitle{Scalable Automatic Differentiation of Multiple
  Parallel Paradigms through Compiler Augmentation}. In
  \bibinfo{booktitle}{\emph{Proceedings of the International Conference on High
  Performance Computing, Networking, Storage and Analysis}} (Dallas, Texas)
  \emph{(\bibinfo{series}{SC '22})}. \bibinfo{publisher}{IEEE Press}, Article
  \bibinfo{articleno}{60}, \bibinfo{numpages}{18}~pages.
\newblock
\showISBNx{9784665454445}
\href{https://doi.org/10.5555/3571885.3571964}{doi:\nolinkurl{10.5555/3571885.3571964}}


\bibitem[Naumann(2012)]%
        {Naumann:2012:TAO}
\bibfield{author}{\bibinfo{person}{Uwe Naumann}.}
  \bibinfo{year}{2012}\natexlab{}.
\newblock \bibinfo{booktitle}{\emph{The Art of Differentiating Computer
  Programs: An Introduction to Algorithmic Differentiation}}.
\newblock \bibinfo{publisher}{Society for Industrial and Applied Mathematics},
  \bibinfo{address}{Philadelphia, PA}.
\newblock
\showISBNx{161197206X}
\href{https://doi.org/10.1137/1.9781611972078}{doi:\nolinkurl{10.1137/1.9781611972078}}


\bibitem[Nocedal and Wright(2006)]%
        {Nocedal:2006:NO}
\bibfield{author}{\bibinfo{person}{Jorge Nocedal} {and}
  \bibinfo{person}{Stephen~J. Wright}.} \bibinfo{year}{2006}\natexlab{}.
\newblock \bibinfo{booktitle}{\emph{Numerical Optimization}
  (\bibinfo{edition}{2} ed.)}.
\newblock \bibinfo{publisher}{Springer New York, NY}.
\newblock
\showISSN{1431-8598}
\href{https://doi.org/10.1007/978-0-387-40065-5}{doi:\nolinkurl{10.1007/978-0-387-40065-5}}


\bibitem[{NVIDIA Corporation}(2025a)]%
        {NVIDIA:2025:CUB}
\bibfield{author}{\bibinfo{person}{{NVIDIA Corporation}}.}
  \bibinfo{year}{2025}\natexlab{a}.
\newblock \bibinfo{title}{{CUB}: Cooperative primitives for {CUDA} {C}++}.
\newblock
\newblock
\shownote{\url{https://nvidia.github.io/cccl/cub/}}.


\bibitem[{NVIDIA Corporation}(2025b)]%
        {NVIDIA:2025:cudss}
\bibfield{author}{\bibinfo{person}{{NVIDIA Corporation}}.}
  \bibinfo{year}{2025}\natexlab{b}.
\newblock \bibinfo{title}{{cuDSS}: Release 0.7.1}.
\newblock
\newblock
\shownote{\url{https://developer.nvidia.com/cudss/}}.


\bibitem[{NVIDIA Corporation}(2025c)]%
        {NVIDIA:2025:thrust}
\bibfield{author}{\bibinfo{person}{{NVIDIA Corporation}}.}
  \bibinfo{year}{2025}\natexlab{c}.
\newblock \bibinfo{title}{{Thrust}}.
\newblock
\newblock
\shownote{\url{https://nvidia.github.io/cccl/thrust/index.html}}.


\bibitem[Provot(1995)]%
        {Provot:1995:DCI}
\bibfield{author}{\bibinfo{person}{Xavier Provot}.}
  \bibinfo{year}{1995}\natexlab{}.
\newblock \showarticletitle{Deformation Constraints in a Mass-Spring Model to
  Describe Rigid Cloth Behaviour}. In \bibinfo{booktitle}{\emph{Proceedings of
  Graphics Interface '95}} (Quebec, Quebec, Canada) \emph{(\bibinfo{series}{GI
  '95})}. \bibinfo{publisher}{Canadian Human-Computer Communications Society},
  \bibinfo{address}{Toronto, Ontario, Canada}, \bibinfo{pages}{147--154}.
\newblock
\showISBNx{0-9695338-4-5}
\showISSN{0713-5424}
\href{https://doi.org/10.20380/GI1995.17}{doi:\nolinkurl{10.20380/GI1995.17}}


\bibitem[Schmidt et~al\mbox{.}(2022)]%
        {Schmidt:2022:TAD}
\bibfield{author}{\bibinfo{person}{P. Schmidt}, \bibinfo{person}{J. Born},
  \bibinfo{person}{D. Bommes}, \bibinfo{person}{M. Campen}, {and}
  \bibinfo{person}{L. Kobbelt}.} \bibinfo{year}{2022}\natexlab{}.
\newblock \showarticletitle{{T}iny{AD}: Automatic Differentiation in Geometry
  Processing Made Simple}.
\newblock \bibinfo{journal}{\emph{Computer Graphics Forum}}
  \bibinfo{volume}{41}, \bibinfo{number}{5} (\bibinfo{date}{Oct.}
  \bibinfo{year}{2022}), \bibinfo{pages}{113--124}.
\newblock
\href{https://doi.org/10.1111/cgf.14607}{doi:\nolinkurl{10.1111/cgf.14607}}


\bibitem[Schreiner et~al\mbox{.}(2004)]%
        {Schreiner:2004:ISM}
\bibfield{author}{\bibinfo{person}{John Schreiner}, \bibinfo{person}{Arul
  Asirvatham}, \bibinfo{person}{Emil Praun}, {and} \bibinfo{person}{Hugues
  Hoppe}.} \bibinfo{year}{2004}\natexlab{}.
\newblock \showarticletitle{Inter-surface mapping}.
\newblock \bibinfo{journal}{\emph{ACM Trans. Graph.}} \bibinfo{volume}{23},
  \bibinfo{number}{3} (\bibinfo{date}{Aug.} \bibinfo{year}{2004}),
  \bibinfo{pages}{870--877}.
\newblock
\showISSN{0730-0301}
\href{https://doi.org/10.1145/1015706.1015812}{doi:\nolinkurl{10.1145/1015706.1015812}}


\bibitem[Sifakis and Barbic(2012)]%
        {Sifakis:2012:FSO}
\bibfield{author}{\bibinfo{person}{Eftychios Sifakis} {and}
  \bibinfo{person}{Jernej Barbic}.} \bibinfo{year}{2012}\natexlab{}.
\newblock \showarticletitle{FEM simulation of 3D deformable solids: a
  practitioner's guide to theory, discretization and model reduction}. In
  \bibinfo{booktitle}{\emph{ACM SIGGRAPH 2012 Courses}} (Los Angeles,
  California) \emph{(\bibinfo{series}{SIGGRAPH '12})}.
  \bibinfo{publisher}{Association for Computing Machinery},
  \bibinfo{address}{New York, NY, USA}, Article \bibinfo{articleno}{20},
  \bibinfo{numpages}{50}~pages.
\newblock
\showISBNx{9781450316781}
\href{https://doi.org/10.1145/2343483.2343501}{doi:\nolinkurl{10.1145/2343483.2343501}}


\bibitem[Smith and Schaefer(2015)]%
        {Smith:2015:BPW}
\bibfield{author}{\bibinfo{person}{Jason Smith} {and} \bibinfo{person}{Scott
  Schaefer}.} \bibinfo{year}{2015}\natexlab{}.
\newblock \showarticletitle{Bijective Parameterization with Free Boundaries}.
\newblock \bibinfo{journal}{\emph{ACM Trans. Graph.}} \bibinfo{volume}{34},
  \bibinfo{number}{4}, Article \bibinfo{articleno}{70} (\bibinfo{date}{July}
  \bibinfo{year}{2015}), \bibinfo{numpages}{9}~pages.
\newblock
\showISSN{0730-0301}
\href{https://doi.org/10.1145/2766947}{doi:\nolinkurl{10.1145/2766947}}


\bibitem[Sorkine and Alexa(2007)]%
        {Sorkine:2007:ARAP}
\bibfield{author}{\bibinfo{person}{Olga Sorkine} {and} \bibinfo{person}{Marc
  Alexa}.} \bibinfo{year}{2007}\natexlab{}.
\newblock \showarticletitle{As-rigid-as-possible surface modeling}. In
  \bibinfo{booktitle}{\emph{Proceedings of the Fifth Eurographics Symposium on
  Geometry Processing}} (Barcelona, Spain) \emph{(\bibinfo{series}{SGP '07})}.
  \bibinfo{publisher}{Eurographics Association}, \bibinfo{address}{Goslar,
  DEU}, \bibinfo{pages}{109--116}.
\newblock
\showISBNx{9783905673463}
\href{https://doi.org/10.5555/1281991.1282006}{doi:\nolinkurl{10.5555/1281991.1282006}}


\bibitem[Tamstorf and Grinspun(2013)]%
        {Tamstorf:2013:DBF}
\bibfield{author}{\bibinfo{person}{Rasmus Tamstorf} {and}
  \bibinfo{person}{Eitan Grinspun}.} \bibinfo{year}{2013}\natexlab{}.
\newblock \showarticletitle{Discrete bending forces and their Jacobians}.
\newblock \bibinfo{journal}{\emph{Graphical Models}} \bibinfo{volume}{75},
  \bibinfo{number}{6} (\bibinfo{year}{2013}), \bibinfo{pages}{362--370}.
\newblock
\showISSN{1524-0703}
\href{https://doi.org/10.1016/j.gmod.2013.07.001}{doi:\nolinkurl{10.1016/j.gmod.2013.07.001}}


\bibitem[Teran et~al\mbox{.}(2005)]%
        {Teran:2005:RQF}
\bibfield{author}{\bibinfo{person}{Joseph Teran}, \bibinfo{person}{Eftychios
  Sifakis}, \bibinfo{person}{Geoffrey Irving}, {and} \bibinfo{person}{Ronald
  Fedkiw}.} \bibinfo{year}{2005}\natexlab{}.
\newblock \showarticletitle{Robust quasistatic finite elements and flesh
  simulation}. In \bibinfo{booktitle}{\emph{Proceedings of the 2005 ACM
  SIGGRAPH/Eurographics Symposium on Computer Animation}} (Los Angeles,
  California) \emph{(\bibinfo{series}{SCA '05})}.
  \bibinfo{publisher}{Association for Computing Machinery},
  \bibinfo{address}{New York, NY, USA}, \bibinfo{pages}{181--190}.
\newblock
\showISBNx{1595931988}
\href{https://doi.org/10.1145/1073368.1073394}{doi:\nolinkurl{10.1145/1073368.1073394}}


\bibitem[Wald et~al\mbox{.}(2026)]%
        {Wald:2026:cuBQL}
\bibfield{author}{\bibinfo{person}{Ingo Wald} {et~al\mbox{.}}}
  \bibinfo{year}{2026}\natexlab{}.
\newblock \bibinfo{title}{{cuBQL}: A CUDA BVH Build-and-Query Library}.
\newblock
\newblock
\shownote{https://github.com/NVIDIA/cuBQL}.


\bibitem[Yu et~al\mbox{.}(2022)]%
        {Chang:2022:MAC}
\bibfield{author}{\bibinfo{person}{Chang Yu}, \bibinfo{person}{Yi Xu},
  \bibinfo{person}{Ye Kuang}, \bibinfo{person}{Yuanming Hu}, {and}
  \bibinfo{person}{Tiantian Liu}.} \bibinfo{year}{2022}\natexlab{}.
\newblock \showarticletitle{MeshTaichi: A Compiler for Efficient Mesh-based
  Operations}.
\newblock \bibinfo{journal}{\emph{ACM Transactions on Graphics}}
  \bibinfo{volume}{41}, \bibinfo{number}{6}, Article \bibinfo{articleno}{252}
  (\bibinfo{date}{Nov.} \bibinfo{year}{2022}), \bibinfo{numpages}{17}~pages.
\newblock
\showISSN{0730-0301}
\href{https://doi.org/10.1145/3550454.3555430}{doi:\nolinkurl{10.1145/3550454.3555430}}


\end{thebibliography}

\clearpage
\appendix

\section{Updating CSR Matrix}
\label{appendix:csr_update}

Algorithm~\ref{alg:csr_insert} shows our procedure for inserting a set of newly introduced nonzero entries (specified in COO format) into an existing sparse matrix stored in CSR format, while preserving the matrix dimensions and updating only its sparsity structure.

\begin{algorithm}
    \caption{Updating CSR matrix by inserting new nonzeros}
    \label{alg:csr_insert}
    \DontPrintSemicolon
    \KwIn{Existing CSR matrix $A$ with row pointer $\texttt{rp}_A[0..R]$, column index $\texttt{ci}_A[0..\mathrm{nnz}_A\!-\!1]$}
    \KwIn{New entries in COO form $(\texttt{nr}[0..S\!-\!1],\texttt{nc}[0..S\!-\!1])$ that are \emph{unique} and not present in $A$}
    \KwIn{Preallocated buffers for output CSR pattern: $\texttt{rp}[0..R]$, $\texttt{ci}[\,]$, and row accumulator $\texttt{acc}[0..R\!-\!1]$}
    \KwOut{Updated CSR pattern $(\texttt{rp},\texttt{ci})$ for $A' = A \cup \{(\texttt{nr},\texttt{nc})\}$}

    \BlankLine
    \textbf{(1) Compute per-row counts.}\;
    \For{$r \gets 0$ \KwTo $R-1$ \textbf{in parallel}}{
    $\texttt{rp}[r] \gets \texttt{rp}_A[r+1] - \texttt{rp}_A[r]$\tcp*{old row nnz count}
    }
    \For{$i \gets 0$ \KwTo $S-1$ \textbf{in parallel}}{
        $\texttt{atomicAdd}(\texttt{rp}[\texttt{nr}[i]], 1)$\tcp*{add 1 for new entries}
    }

    \BlankLine
    \textbf{(2) Prefix sum to build new row pointer.}\;
    $\texttt{rp}[R] \gets 0$\;
    $\texttt{rp} \gets \texttt{ExclusiveScanSum}(\texttt{rp})$\tcp*{length $R\!+\!1$}
    $\mathrm{nnz}_{A'} \gets \texttt{rp}[R]$\;

    \BlankLine
    \textbf{(3) Copy old columns and initialize row offsets.}\;
    \For{$r \gets 0$ \KwTo $R-1$ \textbf{in parallel}}{
    $s \gets \texttt{rp}_A[r]$,\quad $t \gets \texttt{rp}_A[r+1]$\;
    $s' \gets \texttt{rp}[r]$\;
    \For{$j \gets s$ \KwTo $t-1$}{
    $\texttt{ci}[s' + (j-s)] \gets \texttt{ci}_A[j]$\;
    }
    $\texttt{acc}[r] \gets t-s$\tcp*{next free slot within row $r$}
    }

    \BlankLine
    \textbf{(4) Append new columns.}\;
    \For{$i \gets 0$ \KwTo $S-1$ \textbf{in parallel}}{
        $r \gets \texttt{nr}[i]$,\quad $c \gets \texttt{nc}[i]$\;
        $o \gets \texttt{atomicAdd}(\texttt{acc}[r], 1)$\tcp*{row-local offset}
        $\texttt{ci}[\texttt{rp}[r] + o] \gets c$\;
    }

    \BlankLine
    \Return $(\texttt{rp},\texttt{ci})$\;
\end{algorithm}

\FloatBarrier
\newpage

\section{Area-Based Smoothing Code}
\label{appendix:smoothing}
Listing~\ref{list:smoothing_area} shows the full code for how the user expresses the area-based smoothing objective from \S\ref{sec:area} in our system.
\begin{figure}[H]
    \centering
\begin{cppcode}[]{}

using T = float;
T LearningRate = 0.02;  
int NumIter = 20;
Mesh mesh("input.obj");
constexpr int VarDim = 3;
Problem<T, VarDim, VertexHandle> problem(mesh);

// Initialize optimization variables from 
// input positions
auto xInit = mesh.get_input_vertex_coordinates();
problem.var->copy_from(xInit, DEVICE, DEVICE);

// Face-based energy term
problem.add_term<Op::FV>(
  [=] (FaceHandle      fh,    //Face handle
       VertexIterator  iter,  //Face's vertices
       VertexAttribute var) { //Optimization variables
      // The face's vertex positions as 
      // Active variables
      auto x0 = var.active<ActiveT, 3>(fh, iter, 0);
      auto x1 = var.active<ActiveT, 3>(fh, iter, 1);
      auto x2 = var.active<ActiveT, 3>(fh, iter, 2); 
      // Local energy implementation, i.e., face area      
      auto N  = (x1 - x0).cross(x2 - x0);
      return 0.5 * N.norm();});
  
GradientDescent gd(problem, LearningRate);
for (int iter = 0; iter < NumIter; ++iter) {
    problem.eval_terms();
    gd.take_step();
}

\end{cppcode}
\captionof{lstlisting}{Implementation of the area-based smoothing application.}
\label{list:smoothing_area}
\end{figure}

\end{document}